\documentclass[12pt]{article}

\usepackage{bm}
\usepackage{anysize}
\usepackage{amsmath}

\marginsize{2cm}{2cm}{1cm}{4cm}

\date{\vspace{-5ex}}

\begin{document}

\title{The Motion of Small Bodies in Space-time}
\author{Robert Geroch\\ Enrico Fermi Institute\\   The University of Chicago \\ 5640 Ellis
Ave., Chicago, IL, 60637, USA \\ geroch@uchicago.edu
\and
James Owen Weatherall\\ Department of Logic and Philosophy of Science\\
University of California, Irvine \\ 3151 Social Science Plaza A, Irvine, CA,
92697-5100 USA\\ weatherj@uci.edu}

\maketitle

\begin{center}{\bf Abstract}\end{center}
\begin{quote}

We consider the motion of small bodies in general relativity.
The key result captures a sense in which such bodies follow
timelike geodesics (or, in the case of charged bodies, Lorentz-force
curves).  This result clarifies the relationship between 
approaches that model such bodies as distributions supported on
a curve, and those that employ smooth fields supported in small
neighborhoods of a curve.  This result also applies to ``bodies"
constructed from wave packets of Maxwell or Klein-Gordon fields.  
There follows a simple and precise formulation of the optical limit 
for Maxwell fields.

\end{quote}

\begin{center}{\bf \large 1.  Introduction}\end{center}
\vspace{0.5cm}

It has been generally believed, since the work of Einstein and others
\cite{Einstein+Grommer, Einstein+etal}, that 
general relativity predicts,
in some sense, that a small body, free from external interaction, must move
on a geodesic in space-time.  But unraveling the details of that sense has
turned out to be a delicate matter.\footnote{For a review of the 
literature, see \cite{Asada} and \cite{Poisson}. Examples of 
recent work in this area include \cite{Wald+Gralla} and 
\cite{Puetzfeld}.}

Consider a physical theory that incorporates a 4-dimensional manifold 
$M$ of events on which there is specified various fields, subject
to a system of partial differential equations.
Let this system have an initial-value formulation, i.e., be such
that specification of the fields ``initially" determines, by virtue of the
equations, those fields subsequently.  This would seem to be the minimum
requirement for a (non-quantum) physical theory in space-time. Next, let there
be constructed, from these fields, a material body.  Then, by virtue of the
initial-value formulation, every aspect of the future behavior of that body
--- its motion as a whole as well as its detailed internal structure --- is
determined already by those initial conditions. Thus, in a very broad sense,
every physical theory in space-time has the feature that its equations, taken
as a whole, determine the motion of bodies in that theory.   

But the prediction of general relativity is supposed to be more specific: 
That every such body, independent of its composition and construction,
moves, roughly, ``along a geodesic".   How does this come about?
A key feature of general relativity is that all matter must couple to gravity
by means of a stress-energy tensor field, $T^{ab}$, and that this field,
by virtue of the equations on that matter, must be conserved.  In effect,
general relativity singles out, from the full set of equations on
the matter fields, a certain subset.  
The idea is that this subset, which reflects (local) conservation of energy 
and momentum, governs the motion of the body as a whole, while the remaining
matter equations govern the behavior of the internal structure of the body.
It is this universal conservation of stress-energy (together with
a suitable energy condition \cite{Malament}) that is to lead to geodesic 
motion in general relativity.   

Fix an exact solution of Einstein's 
equation, and let there be singled out, in this solution, a ``body".  
For example, in an exact solution representing the sun-earth system, 
we might single out the earth.  We would like to assert that such a
body must be moving along a geodesic.   The problem, of course, is that
we are dealing with an extended body, and it is not clear
what ``along a geodesic" means. 

This problem does not arise, for example, in Newtonian gravity, for there
we have a general result:  The acceleration of the center of mass of an
extended body is given by the mass-weighted average gravitational field
acting on that body.  One strategy would be to try to extend this
Newtonian result to general relativity.  For example, Dixon \cite{Dixon}
defines an
``effective center of mass" trajectory for an extended body in 
general relativity.  The acceleration of that trajectory then serves
as a proxy for the acceleration of the body as a whole.  But, at
least in some cases, this is not a good approximation, for the
acceleration of this trajectory can be strongly affected by the 
local matter distribution in its vicinity --- e.g., by a nearby 
small, massive rock. 

What one would like, ideally, is to produce, for any given 
extended body in general relativity, a number that represents 
``the extent to which that body, taken as a whole, fails 
to follow a geodesic".  Then, one would like to produce a suitable 
upper bound on that number.  Thus, this bound would describe which
features an extended body must possess in order that its motion be
approximately along a geodesic.  But such a general bound seems well 
out of reach.

Here is an alternative strategy.  The first step is to find a context
in which the ``motion of the body", as well as the issue of
whether or not that motion is geodesic, make sense.  That context
involves a body small in size (so we know where it is), and small
in mass (so its mass does not, through Einstein's equation, render
the space-time metric badly behaved).  Such a context  
is the following:  Fix a space-time, $(M, g_{ab})$, together with
a timelike curve $\gamma$ in that space-time.  Here, the space-time
is intended to represent an idealized version of ``the world with the 
body absent"; while the curve represents an idealized ``path of the
body".  Thus, in the case
of the earth, $(M, g_{ab})$ might be the Schwarzschild solution,
and $\gamma$ a curve orbiting the central (solar) mass.  Of course, this
choice is somewhat subjective, for there is no algorithm in general
relativity for constructing ``the same space-time, but with the body 
removed".  The second step is to insert, into this context, some
geometrical object to provide an idealized representation of the 
material content of the body itself.  The idea
is that, insofar as the actual space-time with the actual extended
body resembles this idealized version, then to that extent the 
curve $\gamma$ reflects ``the motion of the actual body".  Absent 
the sort of precise bound described above, we are reduced to such a 
comparison. 

There are two candidates for this geometrical object.   
One, originating with
Matthison \cite{Matthisson} and developed by Souriau \cite{Souriau},
Sternberg and Guillemin \cite{Sternberg+Guillemin}, and others, represents
the body by a certain stress-energy distribution ${\bf T}^{ab}$ with 
support on $\gamma$.  
Provided this distribution is of order zero and has vanishing divergence, 
it follows that $\gamma$ must be a geodesic.  The other 
\cite{Geroch+Jang, Ehlers+Geroch}
represents the body by a family of smooth tensor fields, $T^{ab}$, 
on the space-time.
Provided there exist such $T^{ab}$ that are conserved, satisfy
an energy condition, and are supported in arbitrarily small neighborhoods
of $\gamma$, it again follows that $\gamma$ must be a geodesic. 
The distribution ${\bf T}^{ab}$ or tensor field $T^{ab}$ represents
a sort of idealized limiting stress-energy, i.e., the result of 
scaling the actual stress-energy of a body by a factor that
increases as the mass goes to zero, so arranged to achieve a finite
limit.  

Both of these approaches, then, 
contemplate a limit of a body small in size and mass.  They are
similar in spirit, but quite different in their structure.
Our goal is to connect them.

In Sect 2 we discuss the distributional approach.  We show 
that, for a distribution ${\bf T}^{ab}$ supported on a timelike curve
$\gamma$, conservation and an energy condition already imply that
${\bf T}^{ab}$ must be the stress-energy of a ``point mass", i.e.,
a multiple of $u^a u^b \bm\delta_\gamma$, where $u^a$ is the
unit tangent to $\gamma$ and $\bm\delta_\gamma$ is the delta
distribution of $\gamma$; and
that $\gamma$ must be a geodesic.  A similar result is available
even if conservation is relaxed, i.e., in the presence of external 
forces.  There is a certain freedom, in this case, to incorporate
stresses, originally in ${\bf T}^{ab}$, into the force.  Exploiting
this freedom, there results, for ${\bf T}^{ab}$,
again a point-mass stress-energy; and, for $\gamma$, a curve
satisfying Newton's Law.  We also consider, in Sect 2, 
the case of a particle carrying charge-current, subject to 
electromagnetic forces.  It follows, quite generally, from this distributional
treatment that all electromagnetic moments of the particle of order 
higher than dipole must vanish.  The final equation of 
motion for such a particle is what we expect:  The Lorentz force plus 
certain additional forces arising from the interaction between the 
dipole moments and the gradients of the external field.

In Sect 3, we show how these distributional stress-energies arise as
limits of smooth stress-energies, in the spirit of \cite{Geroch+Jang,
Ehlers+Geroch}.
The main result, Theorem 3, is to the effect that if a collection ${\cal C}$
of smooth fields, $T^{ab}$, conserved and satisfying an energy
condition, tracks sufficiently closely a timelike curve $\gamma$,
then some sequence of fields from that collection actually converges,
up to a factor, to the point-mass distribution; and, therefore,
from conservation, that $\gamma$ must be a geodesic.  The key feature
of this theorem is that we do {\em not} require that the members
of ${\cal C}$ converge to anything at all, but only that
they track, in a suitable sense, the curve $\gamma$.  Then convergence
--- not only to some distribution, but to a specific one ---
follows.  In this sense, then, the distributional
description of particle motion reflects the strategy of 
\cite{Geroch+Jang, Ehlers+Geroch}.
Theorem 3 has a simple and natural generalization to bodies
carrying charge.  

In section 4, we consider a class of examples:  Bodies consisting of 
wave packets constructed from Maxwell or Klein-Gordon fields. 
It turns out that the results of section 3 apply
to such bodies. We thus provide a simple proof that all such packets, in
an appropriate limit, follow geodesics (or, in the case of charged
fields, Lorentz-force curves).  The well-known
``optical limit" of electromagnetic waves is a special case.

\begin{center}{\bf \large Section 2.  Particles}\end{center}
\vspace{0.5cm}

In this section, we consider the motion of a body in the limit
in which that body is confined entirely to a single curve,
i.e., the limit in which the ``path of the body" makes sense.  
We must describe the body, in this limit, in terms of distributions.
A few facts regarding distributions are summarized in Appendices A and B.
Our purpose here is merely to introduce these distributions and describe 
their properties.  The motivation for this treatment appears in Sect 3,
in which we discuss the sense in which the present limit describes the
behavior of actual, i.e., extended, bodies. 

Fix, once and for all, a smooth space-time, $(M, g_{ab})$.

A symmetric tensor $T^{ab}$ at a point of this space-time is said 
to satisfy the {\em dominant energy condition}\footnote{We shall use
the dominant energy condition throughout.  In a substantial fraction
of cases, nothing is changed by replacing this by the
weak, or the strong, energy condition.} at that point
provided:  For any two timelike or null vectors, $u_a, v_b$, at
that point, lying in the same half of the light cone, $T^{ab}u_av_b \geq 0$.  
A symmetric tensor $t_{ab}$ at that point is said to satisfy the
{\em dual energy condition} provided $t_{ab}T^{ab} \geq 0$ 
for every $T^{ab}$ satisfying the dominant energy condition, i.e.,
provided $t_{ab}$ can be written as a sum of symmetrized outer 
products of pairs of timelike or null vectors, all lying in the same 
half of the light cone.  Such a $t_{ab}$ will be called
{\em generic} in case this inequality is strict whenever
$T^{ab}$ is nonzero.  The (closed) cones of tensors satisfying the
dominant energy condition and the dual energy condition are duals of
each other.  The latter cone a proper subset of the former, and
the interior of that cone consists precisely of the generic
$t_{ab}$. 

A symmetric distribution, ${\bf T}^{ab}$, will be
said to satisfy the (dominant) {\em energy condition} provided 
${\bf T}\{{\textsc t}\} \geq 0$ for every test field 
${\textsc t}_{ab}$ satisfying
the dual energy condition everywhere.   

A key fact about distributional stress-energies is the following:
\\

\noindent{\bf Theorem 1.}  Let ${\bf T}^{ab}$ be a symmetric distribution,
satisfying the energy condition.  Then ${\bf T}^{ab}$ is order zero. 
\\

\noindent Proof:  Let ${\textsc t}^1{}_{ab}, {\textsc t}^2{}_{ab}, ... $ 
be a sequence 
of test fields, with
common compact support, $C^0$-converging to test field ${\textsc t}_{ab}$.  
Fix any test field, ${\textsc s}_{ab}$, satisfying the dual energy condition
and generic on the supports of ${\textsc t}_{ab}$ and the 
${\textsc t}^n{}_{ab}$.  Then, 
for every $\epsilon > 0$, both $\epsilon {\textsc s}_{ab} - {\textsc t}_{ab} 
+ {\textsc t}^n{}_{ab}$
and $\epsilon {\textsc s}_{ab} + {\textsc t}_{ab} - {\textsc t}^n{}_{ab}$ satisfy the dual energy
condition
for all sufficiently large $n$.  Applying ${\bf T}^{ab}$ to these two
fields, we conclude: $|{\bf T}({\textsc t}) - {\bf T}({\textsc t}^n)| 
\leq \epsilon 
{\bf T}({\textsc s})$ for all sufficiently large $n$.
\\

In fact, the conclusion of Theorem 1 holds for any distribution 
``arising from tensors whose value, at each point, is restricted
to some proper convex cone".  For
example, it holds also for nonnegative scalar distributions and 
for future-directed timelike vector distributions, as well as for 
symmetric tensor distributions satisfying various other
energy conditions (suitably defined).  

Next, let $\gamma$ be a timelike curve\footnote{We shall take
all curves to be smooth, connected, embedded, without endpoints
and, for timelike curves, parameterized by length.}
on this space-time.
A particle traversing $\gamma$ is represented by its stress-energy:
a nonzero symmetric distribution, ${\bf T}^{ab}$, satisfying
the energy condition and supported on $\gamma$.  
Set ${\bf f}^a = \nabla_b {\bf T}^{ab}$, the (four-)force (density)
that drives ${\bf T}^{ab}$.  So, for example, were there no such
force, ${\bf f}^a = 0$, then the stress-energy would be conserved.
This distribution ${\bf f}^a$ is
also supported on $\gamma$, and is necessarily of order one,
by virtue of the fact that it is expressed as the derivative of
an order-zero distribution.  

Fix any smooth vector field $u^a$ on $M$ that, at points of
$\gamma$, is unit and tangent to this curve.  (Everything will be
independent of how this $u^a$ is extended off $\gamma$.)
We may now decompose ${\bf T}^{ab}$ into its spatial and temporal parts:
\begin{equation}
{\bf T}^{ab} = \bm\mu u^a u^b + 2 u^{(a}\bm\rho^{b)} + \bm\sigma^{ab}.
\label{Tab}\end{equation}
Here, $\bm\mu, \bm\rho^a$ and $\bm\sigma^{ab} = \bm\sigma^{(ab)}$ are 
(unique) order-zero distributions, supported on $\gamma$, and 
having all their indices spatial (i.e., for example,
$\bm\rho^b u_b = 0$).  Here, $\bm\mu$ is interpreted as the mass density
of the particle, $\bm\rho^a$ as an internal momentum density,
and $\bm\sigma^{ab}$ as the stress density.   That the distribution 
$\bm\mu$ is order zero means that this particle can manifest no
higher mass multipole moments:  Indeed, an $n$-pole mass distribution
is of order $n$.  We note that the energy condition
on ${\bf T}^{ab}$ imposes on these three distributions a certain inequality ---
roughly speaking, that $\bm\mu$ be nonnegative and that it bound both 
$\bm\rho^a$ and $\bm\sigma^{ab}$.   

We now turn to the force distribution ${\bf f}^a$.  Its decomposition
is more complicated than that of ${\bf T}^{ab}$, because ${\bf f}^a$ 
is order one rather than zero.  It is convenient to introduce the 
following notion.

A distribution with support on $\gamma$ will be called
{\em local} (to $\gamma$) provided it annihilates every test field that
vanishes on $\gamma$.  For example, every order-zero distribution
with support on $\gamma$ --- such as $\bm\delta_\gamma$ and the delta
distribution of any one point of $\gamma$ --- is automatically local.  
For order-one distributions with support on $\gamma$, some are local 
and some are not.  A distribution of this order is local, roughly speaking,
if it ``takes the derivative (of a test field to which it is 
applied) only in the direction along $\gamma$, 
and not in any orthogonal directions".  Indeed, let $\bm\alpha^a{}_X$ be
any local distribution (where here ``$X$" represents
any arrangement of indices).  Then $\nabla_a \bm\alpha^a{}_X$ is also
local if and only if $\bm\alpha^a{}_X$ is a multiple of the tangent,
$u^a$, to $\gamma$.  Furthermore, every local distribution 
is of this form, i.e., is of the form $\nabla_a (u^a \bm\beta_X)$ for some 
local $\bm\beta_X$.  And, finally, this $\bm\beta_X$ is unique up to 
adding to it $\zeta_X \bm\delta_\gamma$, where $\zeta_X$ is a tensor 
parallel-transported along $\gamma$.   In short, we may ``take the
integral, along $\gamma$", of any local distribution.

The usefulness of the notion of a local distribution
stems from the following fact. 
\\

\noindent{\bf Theorem 2.}  Let ${\bf f}^a$ be any order-one distribution
with support on $\gamma$.  Then
\begin{equation}
{\bf f}^a = \bm\alpha^a + \nabla_b \bm\beta^{ab},
\label{genforce}\end{equation}
where $\bm\alpha^a$ is a local, order-one distribution; and $\bm\beta^{ab}$ 
is an order-zero distribution, spatial in the the index $b$, with support 
on $\gamma$.   Furthermore, the distributions $\bm\alpha^a$ and
$\bm\beta^{ab}$ are unique. 
\\

\noindent Proof.  From the fact that ${\bf f}^a$ is order one, we have:
There exist order-zero distributions $\bm\xi^a$ and $\bm\psi^{ab}$ such that
${\bf f}\{{\textsc t}\} = \bm\xi\{{\textsc t}\} + \bm\psi\{\nabla {\textsc t}\}$
for every test field ${\textsc t}_a$.  
Set $\bm\beta^{ab} = - q^b{}_c \bm\psi^{ac}$ and $\bm\alpha^a = \bm\xi^a
+ \nabla_c(u^c u_d \bm\psi^{ad})$, where $q^b{}_c = \delta^b{}_c + u^b u_c$ 
is the spatial projector.  Eqn. (\ref{genforce}) follows.  Uniqueness
is immediate.
\\

\noindent Eqn. (\ref{genforce}) provides a natural decomposition of 
the force, ${\bf f}^a$, on a particle into a local part (the first 
term on the right) and a
nonlocal part (the second term).  Physically, the distribution
$\bm\alpha^a$ in (\ref{genforce}) represents a total force acting 
on the particle as a whole.  The distribution $\bm\beta^{ab}$, by contrast, 
can be interpreted as describing a ``dipole force":  a pair of equal 
and opposite forces, in the limit in which the magnitudes of those 
forces become large while at the same time the points at which those 
forces act become closer together.  This is, for example, the force
distribution produced by an electric dipole moment in a constant
external electric field.  Theorem 2 can be generalized to 
distributions of higher-order, but not, apparently, to distributions 
with more general support.

Substituting (\ref{Tab}) and (\ref{genforce}) into
the force-equation, $\nabla_b {\bf T}^{ab} = {\bf f}^a$, we obtain
\begin{equation}
\nabla_b \bm\sigma^{ab} + \nabla_b (u^a \bm\rho^b) + \nabla_b (\bm\rho^a u^b)
+ \nabla_b (\bm\mu u^a u^b) = \bm\alpha^a + \nabla_b\bm\beta^{ab}.
\label{gen}\end{equation}
This equation describes the response of the various components
of the stress-energy to a given external force.
We first note that, from uniqueness in Theorem 2, there follows
\begin{equation}
\bm\sigma^{ab} + u^a \bm\rho^b = \bm\beta^{ab}.
\label{stress}\end{equation}
Eqn. (\ref{stress}) implies that the stress of the particle, $\bm\sigma^{ab}$,
as well as its internal momentum, $\bm\rho^a$, are already completely 
determined by the (nonlocal part of the) force.  In physical terms, 
the particle must react to such an external, nonlocal force by adjusting 
its internal stress and momentum to accommodate that force.
It also follows from Eqn. (\ref{stress}) that the spatial projection
of $\bm\beta$ must be symmetric:
\begin{equation}
\bm\beta^{m[b}q^{a]}{}_m = 0.
\label{torque}\end{equation}
Physically, the left side of (\ref{torque}) represents the
torque imposed on the particle by the (nonlocal part of the) external 
force ${\bf f}^a$.  Eqn. (\ref{torque}), then, reflects the
fact that a point particle is unable to absorb torque (storing it internally 
as (spin) angular momentum) for any such storage would violate the 
energy condition.  Were a body to attempt to arrange itself so
as to violate (\ref{torque}),
then it would, in the limit of a point particle, be forced 
to adjust its spatial orientation, instantaneously, in just
such a way that (\ref{torque}) is restored.

As an example of the above, let us consider the special case of a
``free particle":  ${\bf f}^a = 0$, and so ${\bf T}^{ab}$ conserved.  
Then (\ref{stress}) implies that
$\bm\sigma^{ab} = 0$ and $\bm\rho^a = 0$.  Thus, the stress-energy in this
case is given
simply by ${\bf T}^{ab} = \bm\mu u^a u^b$, i.e., is that of a ``mass point".  
Conservation of this stress-energy yields $\nabla_a (\bm\mu u^a) = 0$
and $\bm\mu u^a\nabla_a u^b = 0$.  The former implies that $\bm\mu
= m \bm\delta_\gamma$, where $m$ is a positive number, interpreted as the
mass of the particle, and $\bm\delta_\gamma$ is the delta distribution
of $\gamma$.
The latter implies that the curve $\gamma$ is a geodesic.
Thus, we recover in this special case what we expect:  a point particle, 
of constant mass, moving on a geodesic in space-time.  An important
feature of this example should be noted:  We only impose on the
distribution ${\bf T}^{ab}$ conservation, the energy condition,
and that its support be on $\gamma$ --- but nothing about the form
that ${\bf T}^{ab}$ must take.   From only this input, the specific
form ${\bf T}^{ab} = m u^a u^b \bm\delta_\gamma$ already follows. 

We return now to the general case, with ${\bf f}^a \neq 0$.  We have been
thinking of ${\bf f}^a$ as an effective force, imposed on the particle by
its external environment.   This ``external environment" itself
consists of some additional matter --- possibly of the same type as
that of which the particle is composed, possibly of some different
type.  But when two samples of matter are in interaction with
each other, there is in general no clear-cut way to decide     
how the matter is to be allocated between the two samples.   
Consider, for example, a body carrying a charge distribution, 
placed in a background electromagnetic field.  There results
a total electromagnetic field.  How is the momentum stored in
this total field to be allocated between the body and the background?

This freedom shows itself, in the present instance, in the ability
to add, to ${\bf T}^{ab}$, any symmetric, order-zero distribution 
${\bf s}^{ab}$ with
support on $\gamma$, and simultaneously to add, to ${\bf f}^a$, the divergence
of that distribution.  As noted above, for an ordinary body, of 
finite size, there is in general no natural way to resolve this 
ambiguity between the body and
its environment.  But it turns out that, in the ``particle limit", 
there is such a way.  We choose ${\bf s}^{ab} = -\bm\beta^{ab} 
- u_m \bm\beta^{ma}u^b$.
This ${\bf s}^{ab}$ is symmetric, by virtue of (\ref{stress}); and,
furthermore, results in an order-zero total force.  Indeed, this ${\bf s}^{ab}$
is the most general tensor distribution, constructed from $\bm\beta^{ab}$,
having these properties.  In physical terms, we are allocating any
``dipole force" that may be acting on the particle entirely to the particle.
As a result of this adjustment, we obtain, from (\ref{stress}),  
${\bf T}^{ab} = \bm\mu u^a u^b$.  In short, this adjustment converts
a general ${\bf T}^{ab}$ (given by (\ref{Tab})) and ${\bf f}^a$ 
(given by (\ref{genforce}))
into a new pair having stress-energy that of a
``point particle", subject to a force that is local.  This new pair
is physically equivalent to the original $({\bf T}^{ab}, {\bf f}^a)$:
They describe exactly the same physical situation, but
with a different choice of variables.

The force equation, $\nabla_b {\bf T}^{ab} = {\bf f}^a$, now reduces to
\begin{equation}
\bm\mu A^a = q^a{}_b{\bf f}^b,
\label{mAf}\end{equation}
\begin{equation}
\nabla_b(\bm\mu u^b) = -{\bf f}^b u_b,
\label{masslaw}\end{equation}
where $A^a$ is the acceleration of the curve $\gamma$, and 
$q^a{}_b$ is the spatial projector.  We interpret
Eqn. (\ref{mAf}) as Newton's second law.  Eqn. (\ref{masslaw}) 
represents conservation of mass:  The particle may gain or lose 
mass by virtue of any time-component of the force\footnote{It is
tempting at this point to make a further adjustment between
${\bf T}^{ab}$ and ${\bf f}^a$, so as to achieve ${\bf f}^au_a = 0$.  
Such adjustments exist, but, unfortunately,
there does not appear to be any single, natural one.}.  We emphasize that, in
(\ref{mAf})-(\ref{masslaw}), $\bm\mu$ is simply some non-negative,
order-zero distribution supported on $\gamma$.  It could, for
example, involve $\bm\delta$-distributions of points of $\gamma$.

There is a certain sense in which Eqns. (\ref{mAf})-(\ref{masslaw})
manifest an initial-value formulation.  Imagine, for a moment,
that some force-distribution, ${\bf f}^a$, were specified on the
manifold $M$, once and for all.  We wish to insert a particle
into this environment.  To this end, we choose a
point $p$ of $M$ (the initial position of the particle), a
unit timelike vector $u^a$ at $p$ (the initial 4-velocity of
the particle), and a number $m > 0$ (the initial mass of the
particle).  Then:  The evolution of the point $p$ is determined by
$u^a$; the evolution of $u^a$ is determined (via $A^a$)
by (\ref{mAf}); and the evolution of $m$ is determined by
(\ref{masslaw}). Thus, the
rates of change of these three objects, at each point, are 
determined by the values of these objects at that point.  In other
words, we expect to be able to determine the future 
evolution of the particle.  

But there is a problem with such a formulation, having to do with
the nature of the force itself.  Eqns.
(\ref{mAf}) and (\ref{masslaw}) each assert that one distribution (on $M$) 
is equal to another.  The distributions on the left (arising from the
particle) have support on $\gamma$, and so, therefore, must
the distribution ${\bf f}^a$.  Thus, it makes no sense to
``specify the force, once and for all, as a distribution on $M$". 
Instead, we must specify ${\bf f}^a$ as a distribution {\em with support
on $\gamma$}.  But we cannot do this, within the context of an
initial-value formulation, because we don't know ahead of time
where the curve $\gamma$ will be!  What often happens in specific
examples (as we shall see shortly) is that the particle itself 
is involved in the determination of ${\bf f}^a$, in this manner
achieving an ${\bf f}^a$ supported on $\gamma$.  

There is a curious interaction between (\ref{mAf}) 
and (\ref{masslaw}).  In these equations, the distribution $\bm\mu$ is order
zero, while the distribution ${\bf f}^a$ is permitted to be order one.  
Now, it follows, from the fact that $\bm\mu$ in (\ref{mAf}) is order
zero, that the spatial force, $q^a{}_b {\bf f}^b$, must be zero order.
It is certainly possible to write down a total force, ${\bf f}^a$,
such that its spatial component is order zero, but its temporal
component is genuinely of order one.  But actually to achieve such
a total force, physically, would seem to be a rather delicate
business:  It would mean that the environment, while transferring
mass to the particle in a manner that is genuinely order one, 
at the same time avoids transferring momentum to that same order.  
One would expect that the slightest error in the transfer process would
result in an order-one spatial force, and, as a consequence, in a curve
$\gamma$ that fails to be smooth.   Arguably, such delicate adjustments
are unphysical.  On this basis, then, let us now demand that
the temporal component of the force --- and hence the entire ${\bf f}^a$
--- be order zero.  But now we can repeat the same argument. 
Since now the right side of (\ref{masslaw}) is order zero,
$\mu$ must be of the form
$m \bm\delta_\gamma$, where $m$ is a locally integrable function
on $\gamma$.  But now (\ref{mAf}) implies that the spatial component
of the force have the same character.  On the same physical grounds
as above, we may demand that all of ${\bf f}^a$ be a locally integrable
vector times $\bm\delta_\gamma$.  Repeating this same argument, over
and over, we finally conclude:  Each of $\bm\mu$ and ${\bf f}^a$ must be
of the form of a smooth field along $\gamma$ multiplied by 
$\bm\delta_\gamma$.  In this case, (\ref{mAf})-(\ref{masslaw}) becomes
a simple system of ordinary differential equations along the curve.
We emphasize that the above is merely a rough plausibility argument.

We remark that it is easy to write down explicitly the general 
stress-energy distribution for a (not necessarily free) particle. 
Fix any timelike curve $\gamma$, and any non-negative,
order-zero distribution $\bm\mu$ supported on $\gamma$.  Then
set ${\bf T}^{ab} = \bm\mu u^a u^b$ and ${\bf f}^a = 
\nabla_b (\bm\mu u^a u^b)$.  

We have already noted a special case of (\ref{mAf})-(\ref{masslaw}):
that of a ``free particle" (i.e., that of ${\bf f}^a = 0$).  Then $\bm\mu
= m \bm\delta_\gamma$, where $m> 0$ is a number; and the curve
$\gamma$ is a geodesic.  
Another special case of interest is that of a charged particle.
Let there be given a fixed, smooth, antisymmetric tensor field
$F_{ab}$ on $M$, the Maxwell field.  We think of this $F_{ab}$
as generated by some external charge-current distribution.  Further, 
let our particle manifest its own charge-current distribution ${\bf J}^a$.  
Then ${\bf J}^a$ must be conserved, $\nabla_a {\bf J}^a = 0$, and 
must have support on $\gamma$.  The force ${\bf f}^a$ is then the 
Lorentz force, ${\bf f}^a = F^a{}_b{\bf J}^b$, that $F_{ab}$ imposes
on ${\bf J}^a$.  In order to guarantee that this force be
order one (which is necessary, since ${\bf f}^a$ equals the divergence of
the zero-order distribution ${\bf T}^{ab}$) we demand that this  
charge-current ${\bf J}^a$ also be order one.  Note that there is
no ``self-force" here, i.e., no term involving the interaction of ${\bf J}^a$ 
with the electromagnetic field it produces:  Such an interaction term
would vanish in the present limit.  

We consider first a special case:  We demand that ${\bf J}^a$ actually
be order zero.  It then follows (from uniqueness in Theorem 2)
that ${\bf J}^a = e u^a\bm\delta_\gamma$,
where $e$ is some number, which we interpret as the total charge of
the particle.  Now substitute the resulting Lorentz force into
(\ref{mAf})-(\ref{masslaw}).  It follows from (\ref{masslaw}) that 
$\bm\mu = m \bm\delta_\gamma$, where $m$ is, again, a positive number 
representing the mass of
the particle.  Finally, (\ref{mAf}) implies that the curve
$\gamma$ is a Lorentz-force curve with mass $m$ and charge $e$.
In short, we recover in this special case the standard equation
of motion for a charged point particle.

We turn next to the case of a particle carrying a general charge
distribution --- that in which the charge-current ${\bf J}^a$ is fully of
order 1.  The most general conserved, order-one vector distribution 
${\bf J}^a$ with support on $\gamma$ is given by 
\begin{equation}
{\bf J}^a = eu^a \bm\delta_\gamma + \nabla_b(\bm\tau^{ab}),
\label{J}\end{equation}
where $e$ is a number, and $\bm\tau^{ab}$ is some antisymmetric, order-zero
distribution (not necessarily spatial) with support on $\gamma$. 
To see this, first identify $e$ by applying ${\bf J}$ to test fields given
by the gradient of a function that is constant in a neighborhood of  an 
initial, and also of a final, segment of $\gamma$; and then use the method 
of Theorem 2.  Again, we interpret $e$ as the total electric charge of the particle.
We interpret $\bm\tau^{ab}$ as describing the electric and magnetic
dipole moments of the particle:
\begin{equation}
\bm\tau^{ab} = 2u^{[a}\bm\xi_E{}^{b]} + 
1/2\ \epsilon^{ab}{}_{mn}\bm\xi_B{^mu}^n.
\label{decomtau}\end{equation}
Here, the moments, represented by $\bm\xi_E{}^a$ and $\bm\xi_B{}^a$, 
are order-zero spatial distributions with support on $\gamma$. 
Note that the charge-current 
${\bf J}^a$ can manifest at most dipole --- but no higher --- moments.
The physical reason for this is the following.  Suppose
that the body tried to so arrange its charges to form, e.g., a 
nonzero electric quadrupole moment.  There would result from this arrangement
large electric forces between those charges, which would then
require, in order to hold those charges in place, large stresses.
But large stresses require, by the energy condition, a large mass 
density.  If we now attempt to take the particle limit, retaining
a nonzero electric quadrupole moment, we end up with an infinite
limiting value for the particle's mass.   In short,
higher electromagnetic multipole moments are, in the end, excluded
by the requirement that the particle be described by a well-defined
stress-energy distribution satisfying the energy condition.

Now let the force on our particle be the Lorentz force, $F^a{}_m
{\bf J}^m$, with the charge-current given by (\ref{J}), and decompose this
force as in (\ref{genforce}).  Then the condition (\ref{torque})
becomes
\begin{equation} 
q^{[a}{}_m q^{b]}{}_n (F^m{}_p \bm\tau^{pn}) = 0.
\label{EMtorque}\end{equation}
To interpret this equation physically, decompose the
electromagnetic field into its electric and magnetic parts:
$F_{ab} = 2 E_{[a} u_{b]} + 1/2\ \epsilon_{abmn}B^m u^n$.
Substituting this, and (\ref{decomtau}), we obtain, for the
left side of (\ref{EMtorque}), $E^{[a}\bm\xi_E{}^{b]} + B^{[a}\bm\xi_B{}^{b]}$.
This will be recognized as the physical torque on electric and
magnetic dipole moments placed in an external electromagnetic field.
As a general rule, a body has the freedom to choose its dipole
moments at will as it traverses $\gamma$.  However, in the present
context --- the limit of a point particle --- 
it is necessary that these moments be so chosen that
(\ref{EMtorque}) holds.  Any attempt to do otherwise will result
in a net torque on the particle, which will quickly rotate it so as to 
restore (\ref{EMtorque}).

For the net force on this particle, after incorporating any nonlocal
contributions to the force into the stress-energy, as described earlier,
we obtain:
\begin{equation}
{\bf f}^a = e F^a{}_m u^m \bm\delta_\gamma + \bm\tau^{mn} \nabla_m F_n{}^a
+ \nabla_b[u^b(-F^a{}_n\bm\tau^{nm}u_m + u_nF^n{}_m\bm\tau^{mc} q^a{}_c)].
\label{EMforce}\end{equation}
The right side of (\ref{EMforce}) has a simple physical
interpretation.  The first term is the ordinary Lorentz force on
a point charge $e$.  The second term is the force on 
electric and magnetic dipole moments placed in an external field gradient.
The third term is the time-derivative of a certain vector 
algebraic in the moments and external field.  This vector plays
the role of an effective energy-momentum arising from the interaction 
between the moments and the external field\footnote{The issue of 
what should be called the
``interaction energy-momentum" is somewhat tricky.  An obvious
strategy would be to introduce a (distributonal) solution,
${\bf \tilde{F}}^{ab}$, of Maxwell's equations for charge-current
${\bf J}^a$.  Then take for the ``effective force" the divergence
of the cross-term in the stress-energy of the total electromagnetic 
field ${\bf \tilde{F}}^{ab} + F^{ab}$.  Unfortunately, this strategy will
not work here, for the resulting ``force" will not be local to $\gamma$, 
and furthermore will depend on which solution ${\bf \tilde{F}}^{ab}$ is chosen.
It is perhaps surprising, then, that there turns out to be any natural 
candidate at all for a total electromagnetic force.}.
Eqn. (\ref{EMforce}), then,
requires that any change in this interaction energy-momentum with
time be reflected in a net force on the particle as a whole. 
Note that this total force, given by (\ref{EMforce}), is indeed local, 
and of order one.

In the above, $\bm\tau^{ab}$, which represents the particle's dipole
moments, began as an arbitary order-zero distribution supported
on $\gamma$.  We then found that $\bm\tau^{ab}$ is not so arbitrary,
for it must satisfy the condition, (\ref{EMtorque}) ---that the 
torque on the particle be zero.  There then follows a net total
force on the particle, given by (\ref{EMforce}).  But, as it turns
out, there is a further condition that must be imposed on the
distributon $\bm\tau^{ab}$:  It must be such that the force ${\bf f}^a$
is consistent, via (\ref{mAf}), with a smooth curve $\gamma$.
This condition is somewhat complicated, for the distribution
$\bm\mu$ on the left side of (\ref{mAf}) itself undergoes evolution,
via (\ref{masslaw}).  It turns out, however, that there is at least one
simple way to achieve it:  Let the
distribution $\bm\tau^{ab}$ be given by a smooth tensor along 
$\gamma$, times $\bm\delta_\gamma$.  Then ${\bf f}^a$, given by (\ref{EMforce}),
also takes the form of a smooth vector times $\bm\delta_\gamma$.  
It further follows, from (\ref{masslaw}), that the distribution $\bm\mu$
must be some smooth function $m$ along $\gamma$, times $\bm\delta_\gamma$.
But now the $\bm\delta_\gamma$'s occur as universal factors
in (\ref{mAf})-(\ref{masslaw}), and so may be cancelled out.
Thus, we are left with a simple set of ordinary differential
equations along the curve $\gamma$.

Throughout this section, we have been dealing with 
a particle moving along a timelike curve.  
We now consider the null case.  Thus, let $\gamma$ be a null curve, with
tangent vector $l^a$.  Let ${\bf T}^{ab}$ be a symmetric distribution,
with support on $\gamma$, satisfying the energy condition.  
As before, this ${\bf T}^{ab}$ must be order-zero.  For the case of a null
curve, however, we can no longer decompose tensors into their
spatial and temporal components, and so we cannot write
${\bf T}^{ab}$ as in (\ref{Tab}).  As before, the force driving
this stress-energy is the order-one distribution given by 
${\bf f}^a = \nabla_b {\bf T}^{ab}$.  

Consider first the case of a free particle, ${\bf f}^a = 0$.
Then conservation yields that $\gamma$ is a geodesic, and also that 
${\bf T}^{ab} = \bm\mu l^a l^b$, for some non-negative, order-zero 
distribution $\bm\mu$.
Let us now choose an affine parameter for this geodesic,
and let $l^a$ be the corresponding affine tangent vector.
Then conservation further implies that $\bm\mu = m \bm\delta_\gamma$,
where $m > 0$ is a number and $\bm\delta_\gamma$ is the
delta distribution on $\gamma$ arising from this affine parameterization.
Note that this number $m$ cannot be interpreted as the ``mass", for 
$m$ scales under a change in the choice of affine parameter.

Finally, consider the case of a particle (traveling on a null curve)
subject to a general force.  The notion of a distribution local
to $\gamma$ still makes sense, for $\gamma$ null.  We still have
the formula (\ref{genforce}) for ${\bf f}^a$, but now we cannot require
that $\bm\beta^{ab}$ be ``spatial" in index b.  So, uniqueness fails:  We 
have the freedom to add to $\bm\beta^{ab}$ any distribution
of the form $\bm\zeta^a l^b$, and to $\bm\alpha^a$ the (local) distribution
$-\nabla_b(\bm\zeta^a l^b)$, where $\bm\zeta^a$ is any order-zero
distribution supported on $\gamma$.  We can still carry out the
adjustment, as in the timelike case, to achieve ${\bf T}^{ab} = \bm\mu l^a l^b$
and ${\bf f}^a$ order zero.  But now that adjustment is not unique:
There remains the freedom to move a portion of the distribution
$\bm\mu$ into ${\bf f}^a$.  The force law, $\nabla_b(\bm\mu l^b) l^a + 
\bm\mu A^a = {\bf f}^a$, now requires that ${\bf f}^a l_a$ = 0.  
Non-geodesic null curves are permitted.

\begin{center}{\bf \large 3.  Extended Bodies}\end{center}

In Sect 2, we discussed the motion of a particle --- an idealized body, 
whose path is represented by a curve in space-time and whose 
matter is represented by a certain stress-energy distribution 
having support on that curve.  This treatment turns out to be remarkably simple.
We can write out, explicitly and generally, the distributions representing
the matter as well as any forces that might be acting 
on the particle.  We then determine explicitly the effect of those forces 
on the motion and composition of the particle.  

But this of course is an idealization:  Actual physical bodies have 
finite size.  Our goal in this section is to understand the sense in 
which actual bodies are represented by these idealizations.
Fix a space-time, satisfying Einstein's equation, in which there has 
been identified a ``body".  The general strategy, as discussed 
in Sect. 1, is the following.  First 
write down an idealized space-time and body, 
$(M, g_{ab}, \gamma, {\bf T}^{ab})$,
which in some sense resembles the original system.  We now wish to
compare the actual extended body with its idealization.  To
this end, we introduce some space-times, also satisfying Einstein's
equation, that are intermediate between these two.

To fix ideas, let us consider first the simplest and most manageable
choice of ``intermediate bodies".  First, introduce a family of bodies
whose masses go to zero, while retaining the extended character of the
original body.   The space-time metrics for these bodies approach 
some fixed background metric, $g_{ab}$, on the manifold $M$; while their 
stress-energies go to zero, giving rise to a linearized 
field\footnote{In more detail, we imagine
a one-parameter family, $g(\lambda)_{ab}$, of metrics on $M$, jointly
smooth in $\lambda$ and point of $M$, satisfying Einstein's equation
with stress-energy tensors $T(\lambda)^{ab}$.  
Then set $g_{ab} = g(0)_{ab}$ and $T^{ab} = (d/d\lambda) T(\lambda)^{ab}
|_{\lambda = 0}$.}, $T^{ab}$, on $M$, defined only up to an overall factor.  
This $T^{ab}$ inherits the energy condition from its predecessors. 
Next, allow the sizes of the bodies to go to zero.  Thus, we end 
up with a collection of fields,
$T^{ab}$, on a fixed space-time, which collapse down onto some
timelike curve $\gamma$.

We must introduce a suitable sense of this ``collapsing down".  To this end,
fix a space-time, $(M, g_{ab})$, and a curve $\gamma$ in this
space-time.  Fix also some collection ${\cal C}$ of symmetric tensor fields
$T^{ab}$ on $M$, each of which satisfies the (dominant) energy condition.  
We will say
that this collection {\em tracks} $\gamma$ provided:  Given any test field
${\textsc x}_{ab}$, satisfying the dual energy condition in a neighborhood of
$\gamma$ and generic\footnote{Recall, from Sect 2, that a tensor 
$x_{ab}$, is said to satisfy the dual energy condition at a point provided
$T^{ab}x_{ab} \geq 0$ for every $T^{ab}$ at that point satisfying the 
(dominant) energy condition; and to be {\em generic} at that point 
provided this inequality is strict whenever $T^{ab} \neq 0$.}
at some point of $\gamma$,
there exists an element $T^{ab}$ in the collection ${\cal C}$ such that 
${\bf T}\{{\textsc x}\} > 0$.  This is a key definition.  Note that for
it we impose the energy condition (which plays a crucial role), but not 
conservation.

The idea of this definition is the following.  Let a collection
${\cal C}$ track a timelike curve $\gamma$.  Consider a test field
${\textsc x}_{ab}$ that satisfies the dual energy condition in a narrow
neighborhood of $\gamma$, but then, just outside that neighborhood, goes
quickly to a very large negative multiple of a field satisfying the dual
energy condition.    By tracking, there must be a $T^{ab}$
in ${\cal C}$ with ${\bf T}\{{\textsc x}\} > 0$.  In the integral
that comprises ${\bf T}\{{\textsc x}\}$, that neighborhood
will contribute positively; and the region outside negatively.  Thus,
the bulk of $T^{ab}$ must lie within this narrow neighborhood.
But tracking requires that the collection 
${\cal C}$ contain such a $T^{ab}$ for {\em every}
such test field ${\textsc x}_{ab}$.  In short, tracking means that ${\cal C}$
includes fields $T^{ab}$ the vast majority of whose matter 
clings, as closely as we wish and for as long as we wish, to $\gamma$.

Note that a given collection ${\cal C}$ can track more than one timelike curve.
For example, the collection ${\cal C}$ of {\em all}  $T^{ab}$
satisfying the energy condition in any space-time tracks every timelike
curve in that space-time.  Generally speaking, modifying the $T^{ab}$ 
far from a curve $\gamma$ does not affect whether or not that collection 
tracks $\gamma$.  Thus, a collection ${\cal C}$ could track a
timelike curve $\gamma$ even though {\em every} element of this
collection includes a large amount of matter far from $\gamma$---so long as 
that extraneous matter manages to change its location in $M$, in a suitable
manner (i.e., so as to avoid, eventually, every fixed test field), 
as we go through the various elements of ${\cal C}$.

We consider first the case of ``free bodies", i.e., those described
by stress-energies that are conserved.  Here is the key theorem:
\\

\noindent {\bf Theorem 3.}  Let $(M, g_{ab})$ be a space-time, $\gamma$ a
timelike curve therein, and ${\cal C}$ a collection of fields
$T^{ab}$, each satisfying the (dominant) energy condition, that tracks
$\gamma$.  Let each of these fields be conserved.
Then there exists a sequence, $\overset{1}{T}{}^{ab}, \overset{2}{T}{}^{ab},
\cdots$, each a positive multiple of some element of ${\cal C}$, that
converges, in the sense of distributions, to $u^a u^b {\boldsymbol
\delta}_\gamma $.
\\

\noindent Proof.  First, choose a test field ${\textsc x}^0_{ab}$ that satisfies the dual
energy condition everywhere and is generic at some point of $\gamma$, 
normalized by $(uu\bm\delta_\gamma)\{{\textsc x}^0\} = 1$.
Second, choose a sequence, $\overset{n}{\textsc v}_a$, of test fields such that,
setting $\overset{n}{\textsc x}_{ab} = {\textsc x}^0_{ab}/n 
+ \nabla_{(a}\overset {n}{\textsc v}_{b)}$, each 
$\overset{n}{\textsc x}_{ab}$ satisfies the dual energy condition 
in a neighborhood of $\gamma$, and is generic on some segment $\gamma_n$ of 
$\gamma$, where these segments are increasing and have union 
all of $\gamma$.  [To do this, first fix any extension, $u^a$, of the tangent
to $\gamma$ to a unit timelike vector field, and then set
$\overset{n}{\textsc v}_a$ a function times $u_a$, where this function
is so chosen to achieve the required properties.]
Third, choose a sequence $\overset{1}{\textsc y}_{ab}, 
\overset{2}{\textsc y}_{ab}, 
\cdots$ of test fields, each vanishing on $\gamma$, such that i) each
$\overset{n}{\textsc x}_{ab} - \overset{n}{\textsc y}_{ab}$ satisfies the dual
energy condition in a neighborhood of $\gamma$; and ii) for every
test field ${\textsc m}_{ab}$ that vanishes on $\gamma$, 
$\overset{n}{\textsc x}_{ab} + \overset{n}{\textsc y}_{ab} 
- {\textsc m}_{ab}$ satisfies the dual energy condition everywhere, for
all sufficiently large $n$. [Choose each $\overset{n}{\textsc y}_{ab}$ 
to satisfy
the dual energy condition and be generic wherever it is nonzero.
It rises quickly off the segment $\gamma_n$, and then remains large
in some region away from $\gamma$.  As $n\rightarrow\infty$, the
rate of rise, the size of that region, and the values of 
$\overset{n}{\textsc y}_{ab}$ in that region all increase.]
Finally, for each $n$ choose, by tracking, $\overset{n}{T}^{ab}$, 
a multiple of an element
of ${\cal C}$, such that $\overset{n}{\bf T}\{\overset{n}{\textsc x} 
- \overset{n}{\textsc y}\} > 0$, 
normalized by $\overset{n}{\bf T}\{{\textsc x}^0\} = 1$.  

Now let ${\textsc p}_{ab}$ be any symmetric test field.  Choose test
vector field ${\textsc w}_a$ such that ${\textsc m}_{ab} = 
{\textsc p}_{ab} -
(uu \bm\delta_\gamma)\{\textsc p\} {\textsc x}^0_{ab} 
- \nabla_{(a} {\textsc w}_{b)}$
vanishes on $\gamma$.  
[Here, we make use of the following fact:  For ${\textsc z}_{ab} = 
{\textsc z}_{(ab)}$
any test field satisfying $(uu\bm\delta_\gamma)\{{\textsc z}\} = 0$, 
there exists a
test field ${\textsc w}_a$ such that ${\textsc z}_{ab} - 
\nabla_{(a}{\textsc w}_{b)}$ vanishes on
$\gamma$.] We now have, for all sufficiently large $n$,
\begin{equation}
|(\overset{n}{\bf T}-u u \bm\delta_\gamma)\{{\textsc p}\}| 
= |\overset{n}{\bf T}\{{\textsc m}\}|
\leq \overset{n}{\bf T}\{\overset{n}{\textsc x} + \overset{n}{\textsc y}\} 
\leq \overset{n}{\bf T}\{2\overset{n}{\textsc x}\} = 2/n.
\label{TTproof}\end{equation}
The first step follows from the definition of ${\textsc m}$, the normalization 
of $\overset{n}{T}$, and conservation; the second,
for all sufficiently large $n$, from the defining property of the 
$\overset{n}{\textsc y}$;
the third, from the defining property of $\overset{n}{T}$; and 
the fourth, from the defining property of $\overset{n}{\textsc x}$, 
the normalization of $\overset{n}{T}$, and conservation.  The result follows.
\\

Theorem 3 asserts, in short, that any family of conserved
stress-energies that ``collapse down" onto $\gamma$, in
a suitable sense, necessarily includes a sequence that converges to a 
certain distribution --- that of a ``point particle" --- supported
on $\gamma$.  Note an important feature of the theorem.  We
impose on the family of $T^{ab}$ only conditions reflecting the
locations and the sizes of the bodies they represent, but no conditions
on the form that $T^{ab}$ takes; nor any on its limiting behavior.  
Yet, we conclude from this that some sequence from this family must 
converge, in a suitable sense, to {\em some} distribution, and,
additionally, the specific form of that limiting distribution. 
It is easy to show from this theorem that, if a collection ${\cal C}$
contains a sequence that, possibly after rescaling, converges to 
{\em some} nonzero distribution supported on $\gamma$, then ${\cal C}$
tracks $\gamma$; and furthermore that that distribution is, up to a factor, 
precisely $u^a u^b \bm\delta_\gamma$.   In short, $u^a u^b
\bm\delta_\gamma$ is the unique distribution that arises, under
conservation, from tracking.  Note also that ${\cal C}$ is
presented as merely an unordered (possibly uncountable) collection of 
stress-energies, with no hint as to which of its elements are close 
to the final distribution.  The actual converging sequence is generated
by the theorem.

Here, then, is a sense in which ``small, free bodies in general relativity
traverse geodesics".  First fix the curve $\gamma$.  Then demand that this
curve ``be followed by such bodies", in the sense that there is some
collection ${\cal C}$ of conserved $T^{ab}$ fields, each satisfying the energy
condition, that tracks $\gamma$.  Now apply Theorem 3.  Since all the
$T^{ab}$ in ${\cal C}$ are conserved, so must be the limiting distribution,
$u^a u^b \bm\delta_\gamma$.  But, as we saw 
in Sect 2, conservation of $u^a u^b \bm\delta_\gamma$ implies that 
$\gamma$ is a geodesic.

Here is an example of an application of the theorem.  Fix a timelike 
curve $\gamma$ in a space-time.
Suppose that the following condition were satisfied:  Given any 
compact neighborhood $C$ of any point of $\gamma$, and any
neighborhood $U$ of $\gamma$, there exists a symmetric $T^{ab}$, conserved and 
satisfying the energy condition, that is nonzero somewhere in 
$C \cap U$ and vanishes in $C - U$.  This condition means, in other words, 
that, locally, there exist conserved stress-energies that satisfy the
energy condition and are confined arbitrarily closely to $\gamma$.
It is immediate that the collection ${\cal C}$ of all the $T^{ab}$ 
generated in this way tracks $\gamma$.  Therefore, by
Theorem 3, the curve $\gamma$ must be a geodesic.
This is essentially the result of \cite{Geroch+Jang}.  (To make this comparison
more transparent,  we have replaced the condition on the $T^{ab}$
in \cite{Geroch+Jang} by a ``local" version.)  But there is a 
significant difference
between Theorem 3 and \cite{Geroch+Jang}:  The former, but not the
latter, is applicable to a collection of bodies even if every body
in that collection manifests some (but not too much) matter well
outside of $\gamma$.  This is a useful feature, for, as we shall see 
in the next section, it allows us to apply the theorem
to certain wave packets.

We remark that, if these $\overset{n}{T}{}^{ab}$ are expressed in terms of
various matter fields, then, even though this sequence converges to a
distribution, it need not be true in general that those matter fields 
converge to anything at all --- distributional or otherwise.  Indeed, 
the collection ${\cal C}$ could encompass some (idealized) stars, some 
rocky planets, some pieces of wood, etc.

Consider any collection of bodies, each satisfying the energy
condition, that collapses down to a curve, in the sense of Theorem 3.
Then, according to that theorem, the final limit must be a particle
with zero spin (per unit mass).  Thus, if one wishes treat
"spinning particles" within this framework, then such particles must 
arise either i) from matter violating the energy condition, or 
ii) from a limit different from that envisioned in Theorem 3.  
Neither of these strategies appears attractive.  Compare, 
\cite{Papapetrou}.

Theorem 3 suffers from an apparent defect:  It is global 
on the curve $\gamma$, in the sense that its hypothesis requires
the existence of appropriate bodies along the entirety of this
curve.   Suppose, for example, that we wished to determine
whether the earth, at the present epoch, travels, approximately, on 
a geodesic.  In order to apply Theorem 3, we must introduce a
timelike curve $\gamma$ to represent the earth for all time; and
then assert the existence of $T^{ab}$ that track that curve,
both currently and in the distant past and future.
How are we to know whether
this is possible (or even what $\gamma$ will be) given only the earth
at the present epoch?  This defect, however, is easy to remedy.  
Fix any finite segment of $\gamma$.  Then choose an open neighborhood 
of that segment, 
regard that neighborhood as a space-time in its own right, and apply 
in that space-time Theorem 3.  We thus conclude (having only imposed  
conditions local to that segment) that that segment of $\gamma$ must 
be a geodesic.

The discussion above has been for the case of ``free" bodies, i.e., 
those represented by a conserved stress-energy.  We now consider 
the case of bodies that interact with their environment.   

The simplest case is that of a body carrying charge.  Again, we imagine
a process in which, first, the mass (and charge density) go to zero,
maintaining the extended character of the body; and thereafter the
geometrical size goes to zero.  Thus, we end up
with a fixed space-time, $(M, g_{ab})$, and a fixed 
timelike curve $\gamma$ on that space-time.  On this
space-time, there is specified a fixed background electromagnetic field,
$F_{ab}$, arising from whatever charge-current distribution is present in the
environment. The body itself is described by a pair of fields, $(T^{ab}, J^a)$,
where this pair is defined only up to overall scaling, i.e., up to multiplying
both fields by the same positive factor.  The stress-energy $T^{ab}$ 
must satisfy the
energy condition and the force-law, $\nabla_b T^{ab} = F^a{}_b J^b$; and the
charge-current must satisfy conservation, $\nabla_a J^a = 0$.

Now consider a collection, ${\cal C}$, of such pairs.  Suppose that
the $T's$ of this collection track $\gamma$, as described above.  We
would like to apply Theorem 3 to this situation.  The problem, of
course, is that conservation of the $T^{ab}$, which was used in Theorem 3,
is now replaced by $\nabla_b T^{ab} = F^a{}_b J^b$.  Clearly, we need
to exert some control over the right side of this equation, i.e., over
the charge-current $J^b$.  Indeed, a general charge-current can manifest 
electric and magnetic dipole moments, and, as we saw in Sect 2, these 
moments can affect the motion of a body.  Furthermore,
there is no guarantee that, as the family of bodies collapses down onto
$\gamma$, the dipole moments of these bodies will converge to anything 
at all.  We therefore proceed as follows.

Fix, on a space-time, a pair of fields, $(T^{ab}, J^a)$, the former satisfying
the energy condition.  We say that a nonnegative number $\kappa$
is a bound on the charge-mass ratio for this pair provided:  For
any unit timelike vector $t^a$ at any point of $M$, 
\begin{equation}
 |J^at_a| \leq \kappa T^{ab} t_a t_b.  
\label{chargemass}\end{equation}
In physical terms, this means that, according to
any observer located anywhere in this space-time, the ratio between the 
locally measured charge and mass densities of the material represented by 
$(T^{ab}, J^a)$ is bounded by the number $\kappa$.

This appears to be a reasonable condition to impose on matter.  It
holds, for example, for any material composed (in a suitable sense)
of electrons, neutrons and protons.  It also holds for a charged
fluid, as well as a charged stressed solid, under a suitable additional 
condition on the function of state\footnote{This condition, in the
case of a charged fluid, is the following.  Such a fluid is described
by a 2-manifold of internal fluid states; so the mass and charge
densities, $\mu$ and $\rho$, as well as the pressure $p$, are all
functions on that manifold.  We demand that there exist an open 
subset $U$ of this
2-manifold such that i) the ratio  $\rho/\mu$ is bounded in $U$;
and ii) on the boundary of $U$, the gradient of $p$ 
is tangent to that boundary.  It follows from these conditions that,
given any sample of this fluid with its internal state initially 
lying in $U$, then this sample will so evolve to always remain within $U$,
i.e., will maintain in the course of its evolution a bounded ratio $\rho/\mu$.}.
Further, if the condition above
holds for two types of matter, then it holds when both types are
present (and allowed to overlap) in space-time.  The $\kappa$-value
for this combination is given by the greater of the values for the individual 
matter types.  The condition of bounded charge-mass ratio is easily
generalized to the case in which $T^{ab}$ and $J^a$ are both distributions.
For distributions supported on a curve, as considered in Sect 2,
this condition requires that the dipole and 
higher moments
of $J^a$ all vanish.  Finally, we remark that, if the number $\kappa$ bounds
the charge-mass ratio for $(T^{ab}, J^a)$, then we further have
\begin{equation}
|J^a J^b (g_{ab} + 2t_at_b)|^{1/2} \leq 3 \kappa T^{ab} t_a t_b
\label{currmass}\end{equation}
for every unit timelike $t^a$.
In other words, we have also that the current density of the material 
is bounded by the mass density.  Eqn. (\ref{currmass}) also holds with
the tensor $t_a t_b$ on the right replaced by a suitable multiple of 
any tensor satisfying the generic dual energy condition.  

It turns out that the bound described above is just what is necessary
to generalize Theorem 3 to charged bodies.
\\

\noindent{\bf Theorem 4.}  Let $(M, g_{ab})$ be a space-time, $F_{ab}$ 
an antisymmetric tensor field on $M$, and $\gamma$ a timelike curve.  Let 
${\cal C}$ be a collection of pairs, $(T^{ab}, J^a)$, of tensor fields
on $M$, where $T^{ab}$ satisfies the energy condition, such that each 
satisfies $\nabla_b T^{ab} = F^a{}_b J^b$ and $\nabla_a J^a = 0$; and
each has charge-mass ratio bounded by $\kappa$, where $\kappa \geq 0$
is some fixed number.  Let this collection ${\cal C}$ track $\gamma$.  
Then there exists a number $\kappa'$ satisfying $|\kappa'| \leq \kappa$, 
along with a
sequence of pairs, $(\overset{n}{T}{}^{ab}, \overset{n}{J}{}^a)$, 
each a multiple of some element of ${\cal C}$, that converges to
$(u^a u^b \bm\delta_\gamma, \kappa' u^a \bm\delta_\gamma)$.
\\

The proof consists, first, of repeating the proof of Theorem 3, including,
and suitably bounding, the additional terms arising from the 
electromagnetic interaction.  Conservation of the 
$\overset{n}{T}{}^{ab}$ was used at two points in that proof:  In the
first and fourth steps of Eqn. (\ref{TTproof}).   These two steps
now give rise, in (\ref{TTproof}), to additional terms 
$-(\nabla_b\overset{n}{\bf T}{}^{ab})\{{\textsc w}\}$ and
$-(\nabla_b\overset{n}{\bf T}{}^{ab})\{\overset{n}{\textsc v}\}$, respectively.
For the first term, we have
\begin{equation}
|(\nabla_b\overset{n}{\bf T}{}^{ab})\{{\textsc w}_a\}|
= |\overset{n}{\bf J}{}^b\{F_{ab}{\textsc w}^a\}|
= |\overset{n}{\bf J}{}^b\{{\textsc s}_b\}|
\leq \overset{n}{\bf T}\{\overset{n}{\textsc x} + \overset{n}{\textsc y}\}.
\label{firstterm}\end{equation}
The first step uses the force law.  For the second step, choose (as
we always may) 
${\textsc w}_a$ to be tangent to $\gamma$ on $\gamma$, whence 
$F_{ab}{\textsc w}^a$ is
orthogonal to $u^a$ on $\gamma$.  But every such test field can be written
as a gradient (which is annihilated by $\overset{n}{\bf J}$, by
conservation) plus a test field, ${\textsc s}_b$, that vanishes on $\gamma$.  
The third step, for all sufficiently large $n$ follows from the fact that
${\textsc s}_b$ vanishes on $\gamma$, and that the charge-mass ratio of 
$(\overset{n}{T}{}^{ab}, \overset{n}{J}{}^a)$ is bounded.  
The second term is
converted, in a similar manner, to $\overset{n}{\bf J}\{\overset{n}{\textsc s}\}$,
where $\overset{n}{\textsc s}_b$, again, is a test field vanishing on $\gamma$.
But, in the proof of Theorem 3,  $\overset{n}{\textsc v}_b$, and so this 
$\overset{n}{\textsc s}_b$, is chosen before we must choose 
$\overset{n}{\textsc y}_{ab}$.  
So, using boundedness of the charge-mass ratio, we simply adjust our 
choice of $\overset{n}{\textsc y}$, for each successive $n$, so that  
$\overset{n}{\bf J}\{\overset{n}{\textsc s}\}$ is bounded by, say, 
$(1/10) \overset{n}{\bf T}\{\overset{n}{\textsc x} 
+ \overset{n}{\textsc y}\}$.  
Incorporating these two bounds into (\ref{TTproof}), and making
suitable adjustments in the numerical factors, the proof of Theorem
3 goes through as before.  We conclude:  There exists a sequence,
$(\overset{n}{T}{}^{ab}, \overset{n}{J}{}^a)$, each a multiple of an 
element of ${\cal C}$, such that the $\overset{n}{T}{}^{ab}$ converge 
to the distribution $u^a u^b \bm\delta_\gamma$.  

Next, choose test vector field ${\textsc z}_a$ satisfying
$u^a\bm\delta_\gamma\{{\textsc z}_a\} = 1$.  It follows, from
the bound on the charge-mass ratio and the fact that 
$\overset{n}{T}{}^{ab}\rightarrow u^a u^b \bm\delta_\gamma$,
that the numbers $\overset{n}{\bf J}{}^{a}\{{\textsc z}_a\}$ lie in
a compact set.  Hence, we may, taking a subsequence if necessary,
assume that the $\overset{n}{\bf J}{}^{a}\{{\textsc z}_a\}$ converge
to some number, $\kappa'$.  But every test field is equal
to the sum of a multiple of ${\textsc z}_a$ and a test field, 
${\textsc s}_a$, that satisfies $u^a\bm\delta_\gamma
\{{\textsc s}_a\} = 0$, and, therefore, $\overset{n}{\bf J}{}^{a}
\{{\textsc s}_a\} \rightarrow 0$.  It follows that
$\overset{n}{J}{}^{a} \rightarrow \kappa' u^a\bm\delta_\gamma$.

Thus, under the requirement of bounded charge-mass ratio, the family
${\cal C}$ includes, up to a factor, bodies whose stress-energies
approach that of a point mass while, furthermore, their charge-currents 
approach that of a point charge.
But the $(\overset{n}{T}{}^{ab},
\overset{n}{J}{}^a)$ satisfy the force law, and so therefore,
taking the limit, must
$(u^a u^b \bm\delta_\gamma, \kappa' u^a \bm\delta_\gamma)$.  We
conclude, then, that $\gamma$ must be a Lorentz-force curve, with
charge-mass ratio, $\kappa'$, satisfying $|\kappa'|\leq \kappa$. 

In this sense, then, charged bodies move on Lorentz-force curves.
Again, we emphasize that we do {\em not} require that the
$(T^{ab}, J^a)$ converge to distributions on $\gamma$
--- and certainly not that they converge to any specific distributions.  
Rather, we only
demand that the $(T^{ab}, J^a)$ ``collapse down" onto $\gamma$ in
the sense of tracking; and that, while doing so, they maintain bounded
charge-mass ratio.  It then {\em follows} that these
fields converge to the distributions representing a point mass
and point charge; and, further, that the 
curve $\gamma$ have acceleration appropriate to such a particle.

We remark that the condition, in Theorem 4, that the members of
${\cal C}$ have a uniform bound on the charge-mass ratio, 
can be weakened.  Indeed, all that is actually required in the
proof of 4 is that the $T^{ab}$ bound the $J^a$ ``on average".
This could be expressed, not as pointwise inequalities on these
fields, but rather as inequalities involving the results of applying 
them to certain test fields.

Is there a generalization of Theorem 3 to bodies subject to other forces,
more general than electromagnetic?   Consider a collection
${\cal C}$ of fields $T^{ab}$, each subject only to the energy condition.  
Each of these fields describes a body, where that body is 
subject to an effective
force density, given by $\nabla_b T^{ab}$.  Again, we shall
need to exercise some control over this force.  An obvious condition
is that analogous to Eqn. (\ref{chargemass}) for the charge-current case:
Demand that, for some positive
number $\epsilon$, 
\begin{equation}
|(\nabla_bT^{ab})t_a| \leq \epsilon T^{ab}t_a t_b,
\label{force}\end{equation}
for every unit timelike $t^a$ at every point\footnote{This condition
is also easily generalized to distributions; and, so generalized, 
it implies that, for $T^{ab}$ satisfying the energy condition, 
$\nabla_b T^{ab}$ must be order zero.}.  Here, $1/\epsilon$
represents, in physical terms, a lower limit on the time-scale over which the
external forces can have a significant effect on the body.    

It turns out, however, that this condition alone is not sufficient to
achieve the conclusion of Theorem 3, for the following reason.  
In order to recover Eqn. (\ref{TTproof}), terms involving the
force ($\nabla_b T^{ab}$, applied to certain test vector fields) must be
bounded by terms involving the amount of matter present ($T^{ab}$,
applied to certain test tensor fields).  In the first step 
of Eqn. (\ref{TTproof}) for example, $(\nabla_b\overset{n}{\bf T}{}^{ab})
\{{\textsc w}_a\}$ must be bounded by 
$\overset{n}{\bf T}\{\overset{n}{\textsc x}
+ \overset{n}{\textsc y}\}$.  But here the test field ${\textsc w}_a$ arises
{\em after} we have made our choices of $\overset{n}{\textsc x}$ and 
$\overset{n}{\textsc y}$.  We can always choose ${\textsc w}_a$ to be a multiple of
$u_a$ on $\gamma$, but, even with this further property, no bound of the 
type described in the previous paragraph will suffice.

Fix a unit timelike vector field, $u^a$, that, on $\gamma$, is the
tangent to this curve.  There is a simple physical reason why
merely bounding $u_a\nabla_b T^{ab}$ by 
$T^{ab} u_a u_b$ does not suffice for Theorem 3.  
We may interpret $u_a \nabla_b T^{ab}$
as the rate of mass-transfer (as measured by $u^a$) to the body.
But, if we allow 
mass-transfer to our body, then we cannot expect that the $T^{ab}$ 
must converge to $u^a u^b \bm\delta_\gamma$, for the latter 
represents a particle of {\em constant} mass.   Indeed, we have
already seen these effects, for distributions, in Sect 2.

Such mass-transfer can arise in a variety of contexts.  For example,
for a star passing through a dust cloud, the rest mass of the star
will increase due the accretion of dust.  There are also more
subtle examples.  The act of striking a tennis ball will increase
the rest-mass of that ball, for the stress created by the strike will,
at least in part, be converted into heat within the ball.  Indeed,
it is difficult to think of any scenario in which external forces act
on a body without the possibility, at least in principle, of 
mass-transfer.  

Note that an external electromagnetic field, acting on a body carrying
charge-current, can also result in mass-transfer, by the same mechanism
as for the tennis ball. 
Why, then, did this issue not arise in our earlier treatment 
of charged bodies?  The reason is that in that case we demanded
that the charge-mass ratio of the material remain bounded
--- a very strong requirement.  The mass-transfer
in the electromagnetic case is driven by interaction between the
dipole and higher moments of the body and the external field.  But
the bound on the charge-mass ratio ensures that these moments
(per unit mass) go to zero in the limit, and so too must the
mass-transfer they generate.  This special feature of the electromagnetic
case is reflected in the mathematics as follows.   For ${\textsc w}_a$
a test vector field tangent to $\gamma$ on $\gamma$, the 
effective mass transfer --- the result of applying  
the force density, $F^{ab}J_b$, to that test field --- becomes,
by virtue of conservation, $J^b$ applied to a test field that
{\em vanishes} on $\gamma$.    

There are two possible lines to generalizing Theorem 3 to the case 
of more general forces.  

For the first, we could strengthen the hypothesis of Theorem 3:
We could simply demand that the mass-transfer 
(relative to the amount of matter present) vanish in the limit.  That 
is, we could demand that $(\nabla_b T^{ab})$, applied to any test vector 
field that is tangent to $\gamma$ on $\gamma$, be bounded by $T^{ab}$, 
applied to some test tensor field that vanishes on $\gamma$.  Note that 
this condition has a very different character from those we 
considered above.  Whereas our earlier conditions were imposed on the 
type of matter of which the bodies are composed, this
condition is imposed on the manner in which those bodies 
are constructed.

For the second, we could weaken the conclusion of Theorem 3:  We
could conclude, not that the sequence $\overset{1}{T}{}^{ab}, 
\overset{2}{T}{}^{ab}, \cdots$, converge to the distribution 
$u^a u^b \bm\delta_\gamma$, but rather that, for some sequence
of positive functions $f_n$, the result
of multiplying each $\overset{n}{T}$ by that $f_n$ converge to
this distribution.  This line, in other words, allows mass-transfer,
but adjusts for it by adjusting the $\overset{n}{T}$ before taking
the limit\footnote{There are also lines intermediate between these
two.  We could fix the mass-transfer along $\gamma$, once and for all.
Then, to reflect this choice, we impose
suitable conditions on the $u_a \nabla_b \overset{n}{T}{}^{ab}$ as well as on
the factors by which the $\overset{n}{T}{}^{ab}$ are multiplied.}.

These two lines are perhaps not all that different.  
There is no ``action at a distance" in relativity:  If you wish that
forces be exerted on a body, you must introduce
some other type of matter residing in the immediate vicinity of
that body.   Neither the stress-energy of the matter comprising
the body, nor that of the matter in the environment, will be 
conserved, although of course their sum will be.  It is this
failure of these two types of matter to be separately conserved
that results in a ``force density" on the body.
This scenario requires that the matter that is ``part of the
body" be distinguished from the matter that is ``part of the
environment", and this distinction may not always be clear-cut.
Indeed, the same issue of making this distinction arose in Sect 2.
These two lines, then, merely correspond to different ways of making
this distinction.
Indeed, the whole notion of a body, acted upon by external
forces but otherwise maintaining its integrity, 
is perhaps not as natural in relativity as it is, say, 
in Newtonian mechanics.

Theorem 3 can be generalized to the null case.  To this end,
let $\gamma$ be a null curve in space-time $(M, g_{ab})$, and let
${\cal C}$ be a collection of fields $T^{ab}$ that are conserved, 
satisfy the energy condition, and track $\gamma$.  
Let $t^a$ be a timelike vector field, defined on $\gamma$.  We say that a
choice, $l^a$, of tangent vector to $\gamma$ is {\em affine} 
(with respect to $t^a$) provided $(l^m \nabla_m l^a) t_a = 0$.
It is easy to check that such a tangent vector always exists, and
that it is unique up to multiplication of $l^a$ by a constant.  In the
special case in which $\gamma$ is a (null) geodesic, the affine tangent
vectors (with respect to $t^a$) are the usual geodesic affine tangents,
independent of $t^a$.  So, fix some $t^a$, as well as an affine tangent vector
$l^a$ with respect to that $t^a$.  This choice of $l^a$ generates a
corresponding parameterization of the curve $\gamma$; and, with
respect to that parameterization, $\bm\delta_\gamma$, the delta 
distribution of $\gamma$, makes sense.   We have, for example,
$\nabla_a(l^a \bm\delta_\gamma) = 0$.  Now the proof of Theorem 3 
goes through just as before, with $l^a$ replacing $u^a$ everywhere in that
proof.  For example:  For ${\textsc z}_{ab}$ any symmetric test field along
$\gamma$ satisfying $(l\, l\bm\delta_\gamma)\{{\textsc z}\} = 0$, 
there does indeed
exist a test field ${\textsc v}_a$ such that 
${\textsc z}_{ab} - \nabla_{(a}{\textsc v}_{b)}$
vanishes on $\gamma$ (choosing for ${\textsc v}_a$ a function times $t_a$).  
We conclude that some sequence $\overset{n}{T}{}^{ab}$, multiples of
elements of ${\cal C}$, converge to $l^a l^b \bm\delta_\gamma$.  

Thus, if a collection of conserved $T^{ab}$ satisfying the energy
condition tracks a null curve $\gamma$, then $\gamma$ must be a null
geodesic.  Note that, quite generally,  a collection ${\cal C}$ of 
$T^{ab}$ satisfying the energy condition that tracks every timelike geodesic 
must also track every null geodesic.

In the treatment above, we always
begin with an exact solution of Einstein's equation in which we
have identified some material body; and we always end up with a 
space-time in which there is specified some timelike curve $\gamma$.
Theorem 3 represents just one strategy to get from this
beginning to this end:  Introduce a family of bodies that,
beginning with the given exact solution, have stress-energies
that approach zero, after which the sizes of the bodies also 
approach zero.  

But there are other strategies. 
Fix a space-time, $(M, g_{ab})$, together with a timelike curve
$\gamma$ in this space-time.  Let ${\cal C}$ be a collection of
smooth metrics on $M$, each of whose Einstein tensors satisfies
the energy condition.  Let us now demand:  Given any
neighborhood $U$ of $\gamma$, any compact neighborhood $C$ of a
point of $\gamma$, and any $C^0$-neighborhood of the metric
$g_{ab}$ in $C-U$, there exists a metric $g'_{ab}$ in the
collection ${\cal C}$ such that i) its Einstein tensor is
nonzero somewhere in $C\cap U$ and vanishes in $C-U$; and
ii) the metric $g'_{ab}$, restricted to $C-U$, lies within the given
$C^0$-neighborhood.  This would seem to be the minimal arrangement 
that could be construed as representing a family of bodies that 
``follow, in the limit, a curve
$\gamma$."  The bodies themselves are represented by the metrics in 
the collection ${\cal C}$, and the sense of ``following" is reflected
by condition on the Einstein tensors of these metrics.  Note that 
we impose no conditions whatever on the internal construction of
those bodies (i.e., in $U$):  Their stress-energies can be large 
and can vary rapidly from point to point, and those stress-energies 
can produce large distortions of the space-time metric.  We do, 
however, demand that the external metrics of these bodies approach 
a ``background" in the sense that, in the $C-U$ (i.e., away from
$\gamma$), those metrics $C^0$-approach some fixed metric $g_{ab}$.  

We would not expect to be able to conclude, under this arrangement, 
that $\gamma$ must be a geodesic:  These bodies could,
for example, propel themselves by emitting gravitational
radiation.  Clearly, there is a great deal of room between the
conditions above --- arguably, the weakest possible --- and the very 
strong conditions that underlie Theorem 3.  This suggests the
following program:  Start with the conditions above (which,
apparently, do not restrict the final curve $\gamma$ at all), and then,
in order to conclude that that curve have various properties, 
impose additional conditions on the collection ${\cal C}$. 

Here is an example.  Let us strengthen the conditions above by
demanding that the metrics in ${\cal C}$ $C^1$-converge to $g_{ab}$ 
everywhere --- that is, replace the $C^0$ neighborhood 
of $g_{ab}$ in $C-U$, by a $C^1$ neighborhood of $g_{ab}$ in all of 
$C$.  Clearly, this stronger condition imposes a restriction also on the 
internal structure of the bodies.  Indeed, it amounts, essentially, to the 
requirement that there be a universal upper bound to the their mass 
densities\footnote{A body of mass density $\rho$ and size $L$ 
distorts the metric
by the order of $\rho L^2$; and the derivative operator by
the order of $\rho L$.  These go to zero as $L\rightarrow
0$, provided $\rho$ bounded.}.  Thus, in the example of the earth in orbit
around the sun, this condition contemplates a sequence in
which the earth is replaced successively by a smaller planet, then 
by a rock, then by a grain of sand, etc.   

It turns out that Theorem 3 can be adapted to apply under the 
condition above.  Each metric $g'_{ab}$ in ${\cal C}$ gives rise to
a stress-energy, $T'^{ab}$.  This $T'^{ab}$ is, of course, conserved
with respect to $g'$, but with respect to $g$ it manifests an
effective force.  That force necessarily satisfies Eqn. (\ref{force}),
and furthermore, by $C^1$-convergence, the $g'_{ab} \in {\cal C}$ can 
be so chosen that it further satisfies this equation for 
arbitrarily small $\epsilon$.  Now choose a
sequence $\epsilon_n$ approaching zero sufficiently quickly, and then, 
in Theorem 3, choose each $\overset{n}{T}$ to satisfy (\ref{force}) for 
that $\epsilon_n$.  The bound Eqn. (\ref{force}) suffices to control the 
additional terms $\nabla_a \overset{n}{T}{}^{ab}$ that now arise in 
the first and fourth steps in Eqn. (\ref{TTproof}).
 
We thus conclude, from Theorem 3, that there is a sequence,
$\overset{n}{T}{}^{ab}$, each a multiple of an element of ${\cal C}$, that
approaches the distribution $u^a u^b \bm\delta_\gamma$.  But
approximate conservation of the $\overset{n}{T}$ produces, in the
limit, exact conservation of this distribution.   It follows that
the timelike curve $\gamma$ must be a geodesic.  This is essentially
the result of \cite{Ehlers+Geroch}.

There may be other results along these lines.

\begin{center}{\bf \large 4. Wave Packets}\end{center}

In this section we consider a class of examples, which will serve 
to illustrate the ideas discussed in Sects 2 and 3.  
In general terms, we consider wave packets composed of
solutions of some system of partial differential equations.  We
are interested here in a limit in which the wave packet becomes 
both smaller and longer-lived.  That is, we are interested in a 
limit in which the packet as a whole follows some curve in space-time.

Fix, once and for all, a globally hyperbolic space-time
$(M, g_{ab})$.  We impose global hyperbolicity here solely
in order to guarantee that the solutions of our equations
are sufficient in both number and diversity.  It may be that 
some weaker condition on the space-time will suffice.
Next, consider a system of linear partial
differential equations on some fields on this space-time.
We suppose that we are given an expression for a stress-energy
tensor, $T^{ab}$, quadratic in those fields, and that this 
stress-energy, by virtue of its construction, automatically satisfies 
the energy condition.  We do {\em not} demand that this $T^{ab}$ be 
conserved: There may be external forces acting, through the equations,
on these fields.  Next, fix a timelike or null curve, $\gamma$, in this 
space-time.  Then:  Some given collection, ${\cal S}$, of solutions 
of this system of equations will be said to {\em track} $\gamma$ 
provided the collection of stress-energies, $T^{ab}$, computed from 
those fields tracks $\gamma$ in the sense of Sect 3.

An example of what we have in mind is the Maxwell system.  
Here, we have an antisymmetric tensor field, $F_{ab}$, 
subject to Maxwell's equations (say, with zero sources):  
$\nabla^a F_{ab} = 0$, $\nabla_{[a}F_{bc]} = 0$.
The stress-energy of this field, given by $T^{ab} = F^a{}_mF^{bm} - 1/4
g^{ab}F^{mn}F_{mn}$, satisfies the energy condition and (by
virtue of Maxwell's equations) is
conserved.  Let the collection $\cal{S}$ consist of {\em all} 
solutions of Maxwell's equations in this space-time.   
Which causal curves $\gamma$ does this collection track?  
Since the stress-energy is conserved in this case, it follows
from Sects 2 and 3 that the only candidates for such curves 
are the (timelike or null) geodesics.  It turns out 
that this ${\cal S}$ tracks no timelike geodesics.  
To see this, apply a conformal rescaling to this space-time, i.e.,
replace the metric $g_{ab}$ by $\Omega^2 g_{ab}$, where $\Omega$ 
is some smooth positive function on the manifold.
Every such rescaling preserves Maxwell solutions --- and therefore the 
curves that the collection ${\cal S}$ tracks --- but these rescalings
in general fail to preserve the geodesic character of timelike 
curves.  We conclude, then, that it is only the null geodesics that 
remain as viable candidates for those our collection
${\cal S}$ tracks.

In fact, the collection ${\cal S}$ tracks {\em every} null geodesic
in the space-time $(M, g_{ab})$.  This is most easily seen for
Minkowski space-time.  Fix a null geodesic $\gamma$, and let
${\textsc x}_{ab}$ be a test field that satisfies the dual energy condition
in a neighborhood of $\gamma$ and is generic at some point of
$\gamma$.  Fix a point $p$ of $\gamma$, sufficiently far in the
past along this curve that there is some neighborhood
$U$ of p that does not meet the future of the support of ${\textsc x}_{ab}$.
Then a Maxwell field, generated by initial data supported
in $U \cap I^+(p)$, will meet the support of ${\textsc x}_{ab}$ only in
$I^+(p)$, where $I^+(p)$ denotes the future of $p$.  Furthermore,
the center of mass of this field will be a timelike geodesic passing
through $U$.  Now consider
a sequence of such fields, generated by initial data supported
in successively smaller neighborhoods of $p$, and apply to
these successively larger boosts that preserve both $\gamma$ and $p$.
There results a sequence of Maxwell solutions, each meeting the
support of ${\textsc x}_{ab}$ only in $I^+(p)$, such that their
centers of mass  
converge to $\gamma$.  Clearly, the stress-energies of the solutions 
in this sequence will eventually satisfy ${\bf T}^{ab}\{{\textsc x}_{ab}\} 
> 0$.  That is, this sequence, and so ${\cal C}$ itself, tracks $\gamma$.

This result is easily generalized to curved space-time.  Let $(M, g_{ab})$
be a space-time with Cauchy surface $S$, and let $\gamma$ be any null geodesic  
in this space-time.  Fix a neighborhood $U$ of $\gamma$, a function
$f$ on $M$ having value 1 in some neighborhood of $\gamma$ and
vanishing outside of $U$, and a flat metric,
$g^0_{ab}$, defined in $U$, such that $g_{ab}$ and $g^0_{ab}$,
together with their first derivatives, agree on $\gamma$.
Finally, fix an isometric embedding
of $(U, g^0_{ab})$ in Minkowski space-time, and denote by $\gamma'$
the image of $\gamma$ under this embedding, so $\gamma'$ is also
a null geodesic.
Now, given any solution $F'_{ab}$ of Maxwell's equations in Minkowski 
space-time, set Set $F = F_1 + F_2$,
where $F_1$ is the result of pulling $F'$ back to $U$ via the
embedding and multiplying by $f$; and $F_2$ is the Maxwell field
in $(M, g_{ab})$ that vanishes on $S$ and has sources given
by $(-\nabla_b F_1^{ab}, -\nabla^{[a} F_1^{bc]})$.  Then this $F_{ab}$
is a source-free solution of Maxwell's equation in $(M, g_{ab})$.  
Suppose, next, that
the $F'$ track $\gamma'$ in the Minkowski space-time.  Then, we
claim, the corresponding $F$'s track $\gamma$ in $(M, g_{ab})$.
Indeed, the $F_1$ clearly track $\gamma$.  But we also have a bound
on the sources for $F_2$, as follows from the fact the $F'$ satisfy 
Maxwell's equations in the Minkowski space-time and track $\gamma'$
there, together with the defining properties of $f$ and $g^0_{ab}$.
It follows from this bound that the contribution of $F_2$,
relative to $F_1$, can be made as small as we wish. 

We conclude, then, that the collection ${\cal S}$ of all solutions
of Maxwell's equations in any globally hyperbolic space-time tracks every null
geodesic $\gamma$ in that space-time --- these curves and no other curves.  
This conclusion reflects what is usually called the ``optical limit"
of electromagnetism.  We remark that the present formulation of
the optical limit is precise and remarkably simple.  The key 
idea that makes this happen is the notion of tracking.

We turn next to a second example --- the Klein-Gordon equation.  Fix a
positive number $m$.  Then the Klein-Gordon field is a complex scalar
field $\phi$ on $M$, subject to the equation $\nabla^2 \phi - m^2 \phi = 0$.  
The stress-energy of this field is given by
\begin{equation}
T^{ab} = \nabla^{(a} \phi \nabla^{b)} \overline{\phi}
- (1/2) (\nabla^n \phi \nabla_n \overline{\phi}) g^{ab}
- (1/2) m^2 \phi\overline{\phi} g^{ab}.  
\label{KGTab}\end{equation}
This $T^{ab}$ satisfies the energy condition, and, again, is conserved.  
Again, we let ${\cal S}$ be 
the collection of {\em all} solutions of the Klein-Gordon equation 
(for this fixed value of $m$) in our space-time $(M, g_{ab})$, and 
again we ask for those curves $\gamma$ that this collection tracks.  
It follows, again from conservation, that the only candidates are
the timelike and null geodesics.

We first note that, in the case $(M, g_{ab})$ Minkowski space-time,
 the collection ${\cal S}$ does in fact track every 
null geodesic, by the same argument as for the Maxwell 
case.  This is what we would have expected.  Think of a Klein-Gordon
wave packet as representing a massive particle.  In the high-energy
limit, such a particle would nearly follow a null geodesic; and so 
we expect that the corresponding wave packets would track those curves.

But massive particles at lower energies typically follow timelike geodesics.  
Does the collection ${\cal S}$ track these curves, too?  It turns out 
that it does not.  To see this, suppose,
for contradiction, that ${\cal S}$ did track some timelike geodesic,
$\gamma$.  Then, by Theorem 3, some sequence of stress-energies,
(\ref{KGTab}), must converge, as distributions, to $u^a u^b
\bm\delta_\gamma$.  But this in turn requires, from Eqn. (\ref{KGTab}), 
that $\nabla^{(a} \phi \nabla^{b)} \overline{\phi}$
and $\nabla^n \phi \nabla_n \overline{\phi} + m^2 |\phi|^2$ converge to
$u^a u^b \bm\delta_\gamma$ and zero, respectively.  It follows that
$|\phi|^2$ must converge to $\bm\delta_\gamma/m^2$; and, therefore,
that $\nabla^2(|\phi|^2)$ must converge to $\nabla^2(\bm\delta_\gamma/m^2)$.  
But the former is equal to $2\nabla^n\phi \nabla_n\overline{\phi}
+ 2m^2 |\phi|^2$, which, as we have just seen, converges to zero.  
We now have a contradiction, for $0 \neq \nabla^2(\bm\delta_\gamma/m^2)$.
In short, the Klein-Gordon stress-energy (\ref{KGTab}) has the 
wrong ``shape" to give rise, in the limit, to a point-particle mass 
distribution\footnote{We remark that a similar argument gives an
alternative proof that the solutions of Maxwell's equations
track no timelike curve.}.

This conclusion is what we would expect geometrically.  
Think of a Klein-Gordon wave packet in Minkowski space-time as
composed of plane waves, each with a frequency-wave number vector, 
$k^a$, satisfying $k^a k_a = -m^2$.  In order that such a packet be 
long-lived, it must be the case that the bulk of the waves comprising 
that packet have $k$-values close to some fixed vector, $k^a_0$.  And, 
in order that the packet itself be small in size, it must be the case that
the wavelength associated with this $k_0$ be small.  But a 
frequency-wave number vector $k_0$, of fixed norm, can reflect
small wavelengths only if it lies near the light cone.  In physical terms,
the parameter $m$ that appears in the Klein-Gordon equation is related
to the physical mass by a factor of Planck's constant, $\hbar$.  
It is only in the classical limit, $\hbar \rightarrow 0$, that
we expect wave packets to track curves in the spacetime.  But
this limit corresponds, for fixed physical mass, to $m\rightarrow \infty$.

It is clear, from the discussion above, that Klein-Gordon solutions 
become more and more efficient at forming wave packets
as the Klein-Gordon mass, $m$, of those solutions
increases.  This observation suggests that we proceed as follows.
Let the collection ${\cal S}$ consist, not of the Klein-Gordon solutions
for some fixed value of the parameter $m$, but rather of all solutions 
for {\em all} values of this parameter.  

This collection ${\cal S}$, it turns out, does indeed track every 
timelike geodesic.  To see this, consider first the case in which
$(M, g_{ab})$ is Minkowski space-time.  Fix a ``time function" $t$ in
this space-time, i.e., such that $u^a = \nabla^a t$ is constant,
unit, and timelike.  For each value of $t$, denote by $S_t$ 
the spacelike 3-surface of constant $t$.  For $\zeta$ any smooth, 
complex-valued function on $M$, denote by $E(\zeta, t)$ the value 
of the integral of $T_\zeta{}^{ab}u_b$ over the surface $S_t$, 
where $T_\zeta{}^{ab}$ denotes the result 
of replacing $\phi$ by $\zeta$ in Eqn. (\ref{KGTab}).  Then $E(\zeta,t) \geq 0$;
and we have, by direct computation,  
$\nabla_b T_\zeta{}^{ab} = 1/2 [\nabla^a\zeta
(\nabla^2 - m^2)\overline{\zeta} + \nabla^a \overline{\zeta}
(\nabla^2 - m^2)\zeta]$.  It follows that 
\begin{equation}
dE(\zeta,t)/dt \leq 2 E(\zeta, t)^{1/2}
(\int_{S_t} |(\nabla^2-m^2)\zeta|^2 dS)^{1/2}.  
\label{zeta}\end{equation}
Think of $E(\zeta, t)$ as an ``effective energy" of the function $\zeta$,
and of Eqn. (\ref{zeta}) as reflecting the idea that this energy can
grow only to the extent that the function $\zeta$ fails to satisfy
the Klein-Gordon equation.

Next, fix any static (with respect to $u^a$), complex-valued function 
$\alpha$, of compact spatial support, on the manifold $M$.  
Denote by $\phi$ the solution of the Klein-Gordon equation, for
some value of $m>0$, whose initial data, on the surface $S_0$, 
are $(\alpha|_{S_0}, im\alpha|_{S_0})$; and 
by $\phi'$ the function $\alpha \exp(imt)$ on $M$.  The two 
functions $\phi$ and $\phi'$ manifest the same initial conditions 
on $S_0$, and furthermore their difference satisfies 
$(\nabla^2- m^2)(\phi-\phi') = \exp(imt)\nabla^2\alpha$.  Setting 
$\zeta = \phi - \phi'$ in Eqn. (\ref{zeta}), there follows a bound on 
$E(\phi-\phi',t)$, independent of $m$.  
In the region of $S_t$ outside of the support of $\alpha$, $\phi'$
vanishes; and so this bound also serves as a bound, still independent
of $m$, for the energy-integral of $\phi$, taken only over this region.
But the total energy of $\phi$ is given by $E(\phi, t)
= (1/2)\int_{S_0} (m^2 \alpha\overline{\alpha} +  \nabla^a\alpha
\nabla_a \overline{\alpha}) dA$, which is independent of $t$,
and grows without bound as 
$m\rightarrow \infty$.  We conclude:  The value of the energy
integral of $\phi$, taken over the entire surface $S_t$ dominates
the value of that energy taken only outside the support of
$\alpha$ --- and the extent of this domination increases as $m$ 
increases.  It follows that the collection of Klein-Gordon solutions
$\phi$, constructed as above (for all $m>0$), track every timelike 
geodesic orthogonal to the $S_t$.  Similarly
for other timelike geodesics in Minkowski space-time.

This result is easily generalized to curved space-times, by an
argument similar to that for the Maxwell case.  
We conclude, then, that, in any globaly hyperbolic space-time,
the collection ${\cal S}$ of all Klein-Gordon
solutions for all values of $m > 0$ tracks precisely the timelike
and null geodesics in that space-time.

We next turn to the case of the charged Klein-Gordon field.  Fix,
on the space-time $(M, g_{ab})$, an antisymmetric tensor field $F_{ab}$,
the background field generated by some external sources. 
Fix also  numbers $m$ and $e$.  Then a Klein-Gordon field (for these
values of $m$ and $e$) is a charge-$e$ scalar field $\phi$ (necessarily
complex), subject to the equation $\nabla^2 \phi - m^2 \phi = 0$,
where $\nabla_a$ is the charge-derivative operator.  The stress-energy
of this field (which, again, satisfies the energy condition) is 
given by (\ref{KGTab}), and the charge-current by
\begin{equation}
 J^a = (ie/2)(\overline{\phi}\nabla^a\phi - \phi\nabla^a\overline{\phi}).
\label{KGJ}\end{equation}
There follows:  $\nabla_b J^b = 0$ (charge conservation) and 
$\nabla_b T^{ab} = F^a{}_b J^b$ (force equation).
Thus, the stress-energy in this case fails to be conserved:  There 
is an external
force, the Lorentz force, acting on the charged fields.
Note that, for $\phi$ a charged Klein-Gordon solution, for some 
values $(e, m)$, then $\overline{\phi}$ is also a solution, 
for $(-e, m)$; and these two solutions have the same stress-energy
and charge-current.
It is immediate from (\ref{KGJ}) and (\ref{KGTab}) that the 
charge-mass ratio for a charged Klein-Gordon field is bounded, in the
sense of Sect 3, by the number $|e|/m$.  

Fix, once and for all, a globally hyperbolic space-time, $(M, g_{ab})$,
and a number $\kappa$.  Denote by ${\cal S}_\kappa$ the collection
of all Klein-Gordon solutions on this space-time with $m > 0$
and $e/m = \kappa$.  This collection includes solutions for
arbitrarily large values of $m$:  As we have seen in the
uncharged case, it is only in the limit $m \rightarrow \infty$ that
interesting tracking behavior emerges.  Let $\gamma$ be a timelike
curve in this space-time.

Suppose, in the first instance, that the collection ${\cal S}_\kappa$
tracks $\gamma$.  It then follows, from Theorem 4, that $\gamma$
must be a Lorentz-force curve, with charge-mass ratio in the
closed interval $[-\kappa, \kappa]$.

Next, let there be given a curve $\gamma$, a Lorentz-force curve
with charge-mass ratio $\kappa$.  Then, we claim, the collection 
${\cal S}_\kappa$ tracks $\gamma$.  This follows by essentially
the same argument as in the uncharged case.  (Use a choice of
vector potential for the Maxwell field $F_{ab}$, to convert charged
scalar fields to ordinary (complex) scalar fields; and the charged
derivative operator to the ordinary derivative operator, corrected
by a term involving the vector potential.)  If, in that argument,
we replace $im$ by $-im$, we obtain wave packets that track the
Lorentz-force curves with charge-mass ratio $-\kappa$.  (Indeed,
it is easy to prove, quite generally, that $S_\kappa$ and 
$S_{-\kappa}$ track precisely the same curves, as a consequence of
the fact that the operation of complex conjugation, 
which sends $S_\kappa$ to $S_{-\kappa}$, preserves tracking.)

To summarize, we have shown that every timelike curve tracked by the 
collection ${\cal S}_\kappa$ of charged Klein-Gordon fields is
a Lorentz-force curve with charge-mass ratio in the interval
$[-\kappa, \kappa]$; and, conversely, that every curve whose
charge-mass ratio lies at either endpoint of this interval is, 
indeed, tracked
by ${\cal S}_\kappa$.  Are the Lorentz-force curves with charge-mass
ratio in the interior of this interval also tracked?  We suspect
that they are not, for the following reason.  Fix, say, in
Minkowski space-time, an initial surface $S_0$.  Then
initial data on $S_0$ of the form $(\alpha, im \alpha)$, where $\alpha$
is any function on $S_0$, give rise, as $m\rightarrow \infty$, to
a packet that follows a Lorentz-force curve with charge-mass
ratio $\kappa$; while initial data of the form $(\alpha, -im\alpha)$
gives rise to a packet that follows a curve with ratio $-\kappa$.
But {\em every} set of initial data on $S_0$ can be written,
uniquely, as the sum of one set data of the first form and one
of the second. 

The general conclusion, in any case, is that charged Klein-Gordon fields,
in an appropriate limit, follow the corresponding Lorentz-force
curves.  It is curious that quantum field theory exploits
an entirely different mechanism to achieve Lorentz-force motion.
Consider charged, spin-zero particles in Minkowski space-time.
The Hilbert space of one-particle states is formed, not from
the charged-field solutions of the Klein-Gordon equation, 
but rather from the charge-zero solutions.  The effect of an
external electromagnetic field is represented by an interaction
on this Hilbert space, and it is this interaction that, 
in the classical limit, is responsible for Lorentz-force motion. 

\begin{center}{\bf \large Section 5.  Conclusion}\end{center}
\vspace{0.5cm}

The theory of a ``point particle" --- represented by a distribution
${\bf T}^{ab}$ supported on a timelike curve $\gamma$ --- is
remarkably simple.  Here is a context in which the 
motion of a body makes sense.  And, indeed, we recover, in
this context, what we expect:  that $\gamma$ is a geodesic (in the 
case of no external force),
or a Lorentz-force curve (for a particle carrying charge).
We have argued that this model is a reliable indicator of 
how actual, extended bodies will behave:
If we demand of a body only that it
be ``sufficiently small", in both size and mass, then that
body is already well-represented by the corresponding distribution.
The key notion here is that of tracking.  Indeed, tracking
applies even to ``bodies" constructed as wave packets of, e.g., Maxwell
or Klein-Gordon fields.  This leads, among other things, to
a simple, transparent version of the optical limit for electromagnetism. 

There remain a number of open issues.  What is the mechanism by 
which a particle
achieves the torque-condition, (\ref{torque}) (or, in the electromagnetic
case, (\ref{EMtorque}))?  How, for example, does this condition
emerge from the dynamics of extended bodies?  For a
particle subject to an external force, one can exploit the freedom 
to exchange matter between the particle and its environment to simplify
the stress-energy, leading to Eqns. (\ref{mAf})-(\ref{masslaw}).  
Is there a similar freedom for an extended body; and how does 
it operate in the limit?  Also, we gave
a physical argument that these equations should require that the relevant
distributions --- $\bm\mu$ and ${\bf f}^a$ --- be multiples of 
$\bm\delta_\gamma$.  Can this argument be placed on firmer footing?  
Finally, we noted that Eqns. (\ref{mAf})-(\ref{masslaw}), while they do 
not have a meaningful initial-value
formulation in general, do have such a formulation in the
electromagnetic case.  What is the status of this issue for other forces?
We saw that a particle carrying charge cannot
manifest electromagnetic multipole moments higher than dipole, resulting
in Eqn. (\ref{J}).  Could one see in more detail how this happens, by
considering a limit of extended bodies?  Can Theorem 3 be generalized 
to forces other
than electromagnetic (e.g., contact forces)?   Can it be generalized 
to include other strategies by which particles emerge as limits of 
extended bodies?

Here is a curious example of a further application of the notion
of tracking.  We would certainly expect that general relativity will,
in some sense, prohibit ``tachyonic bodies", i.e., those that follow 
spacelike curves.  Tracking, it turns out, provides a precise 
formulation of this idea.  We claim:
In any space-time, the collection ${\cal C}$ of all conserved $T^{ab}$ 
satisfying the dominant energy condition tracks no spacelike curve.
To see this, suppose, for contradiction, that ${\cal C}$ tracked
some spacelike curve $\gamma$.   Let ${\textsc v}_a$ be a test vector 
field such that $\nabla_{(a}{\textsc v}_{b)}$ satisfies 
the dual energy condition in a neighborhood of $\gamma$ and is 
generic at some point of $\gamma$.  [To construct such a field, fix, in a
neighborhood of $\gamma$, a time function $t$ that is positive on
all but a finite segment of $\gamma$, and set, in that neighborhood,
${\textsc v}_a = f(t) \nabla_a t$, where $f$ is a suitable smooth,
nonnegative function of one variable that vanishes for $t \geq 0$.]
We now have a contradiction, for ${\bf T}^{ab}
\{\nabla_{(a}{\textsc v}_{b)}\}$ must vanish for all $T^{ab}$ in
${\cal C}$, by conservation; but must be positive for some such
$T^{ab}$, by tracking.  It turns out that the dominant energy condition
is essential for this argument.  Indeed, in Minkowski space-time, for
example, the collection of conserved $T^{ab}$ satisfying the weak 
energy condition tracks (with that definition suitably adapted
to that energy condition) every spacelike geodesic.  This follows,
noting that every $T^{ab} = \mu x^a x^b$, where $x^a$ is a constant 
spacelike vector field and the function $\mu$ is nonnegative and 
constant along the $x$-trajectories, is conserved and satisfies 
the weak energy condition.

It is of some interest to ask which of the present results hold also
for Newtonian gravitation.  Recall that, in that theory, there is
specified a 4-manifold of events, on which there is given (among other 
things) a function $t$
(the ``Newtonian time") and a derivative operator $\nabla_a$ (which
includes the effects of gravitation).  Matter, as in relativity,
is described by a conserved, symmetric stress-mass-momentum tensor, $T^{ab}$.
The natural candidate for an energy condition on such a tensor
is the requirement that the combination $T^{ab} \nabla_a t \nabla_b t$ 
(the ``mass density") be nonnegative.
This condition differs in an important way from the
dominant energy condition in relativity:  Whereas the latter 
(which involves $T^{ab}$ contracted with a variety of vectors) 
controls the entire stress-energy tensor, the former (since $T^{ab}$ is 
contracted only with the single vector $\nabla_a t$) leaves
certain components of $T^{ab}$ unrestricted.

Consider, in a Newtonian space-time, a curve $\gamma$, parameterized
by $t$, and a conserved distributional ${\bf T}^{ab}$, satisfying this energy
condition and supported on $\gamma$.  Then it is easy to prove that
${\bf T}^{ab} \nabla_a t \nabla_b t$ must be a multiple of $\bm\delta_\gamma$; 
and that, if this multiple is nonzero, then $\gamma$ must be a geodesic.  
But (in contrast to the case in relativity) the remaining components of
the distribution ${\bf T}^{ab}$ remain essentially free, and, indeed, can be
of arbitrarily high order.   There is no natural generalization of 
the full Theorem 3 to Newtonian gravitation, because the energy condition in 
that theory permits mass to be transported, arbitrarily rapidly, 
from place to place.  There is, however, one result \cite{Weatherall} 
along the lines of Theorem 3:  Given any curve $\gamma$ tracked by a
collection ${\cal C}$ of fields $T^{ab}$ (under a slightly stronger energy
condition than that above, and under a further condition on the
supports of the $T^{ab}$), then that curve must be a geodesic.

In Sect 4, we considered wave packets constructed of Maxwell and
Klein-Gordon fields, and determined which curves these fields
track.  To what extent can these results be extended
to more general systems of equations?  

Consider, for example, Yang-Mills fields.  In a fixed, globally
hyperbolic background space-time, let ${\cal S}$ denote the
collection of Yang-Mills fields with zero source.  
Then tracking makes sense (since the Yang-Mills stress-energy 
satisfies the energy condition); and any timelike or null curve 
this ${\cal S}$ tracks must be a geodesic (since this stress-energy 
is conserved).  We would expect that it
is precisely the null geodesics that are tracked, although the proof
will be considerably more difficult than in the Maxwell case, because
the Yang-Mills equation is nonlinear.  Next, fix a background Yang-Mills
field in this space-time, and consider mass-$m$ boson fields carrying
Yang-Mills charge.  The stress-energy of such a field satisfies the
energy condition, as well as a force-equation involving the Yang-Mills
current of that field.  We would expect behavior similar to that of 
Klein-Gordon:
That the collection of such fields, for fixed $m$, tracks the null 
geodesics; and that that collection, for all $m>0$, tracks also those 
timelike curves manifesting Yang-Mills force.

Consider, as a second example, the Dirac system.
Recall that the Dirac field (of charge
$e$, mass $m$) is given by a pair, $(\xi^A, \eta^{A'})$, of charge-$e$
spinor fields, satisfying
\begin{equation}
\nabla_{AA'}\xi^A = (m/\sqrt{2})\eta_{A'},\ \ \  \nabla_{AA'} \eta^{A'}
= (m/\sqrt{2}) \xi_A.
\label{DiracEq}\end{equation}
The stress-energy and charge-current of this Dirac field are given
by
\begin{equation}
T^{ab} = (i/2)(\xi^A\nabla^b \overline{\xi}{}^{A'}
  - \overline{\xi}{}^{A'}\nabla^b \xi^A
  + \eta^{A'}\nabla^b\overline{\eta}{}^A
  - \overline{\eta}{}^A\nabla^b\eta^{A'}
  + (a\leftrightarrow b)),
\label{DiracT}\end{equation}\begin{equation}
 J^a = e(\xi^A\overline{\xi}{}^{A'} + \eta^{A'}\overline{\eta}{}^A).
\label{DiracJ}\end{equation}
where ``$a\leftrightarrow b$" means the same expression, but with the
roles of ``$a$" and ``$b$" reversed.

Note that for the Dirac field, in
contrast to the Klein-Gordon field, the stress-energy does
not in general satisfy the energy condition\footnote{The easiest
way to see this is to note that there is a complex-conjugation
operation on Dirac fields, similar to that on Klein-Gordon fields,
but in this case it reverses the signs of both the stress-energy and the
charge-current.  Restricting consideration to positive-frequency
solutions of the Dirac equation in Minkowski space-time will not
work, for even this restricted class of solutions fails to
satisfy the energy condition.}.  Consequently, we do not have the notion of
a family of Dirac solutions tracking a curve, in the sense of
Sect 3.  

In any case, we might expect that there can be constructed,
from solutions of the Dirac equation, wave packets (suitably defined)
that follow timelike curves.  Fix a number
$\kappa$.  Then, we might expect that, from the collection of
all solutions of the Dirac equation with charge-mass ratio
$\kappa$, there will be a sequence that follows (in some
suitable sense) any Lorentz-force curve with charge-mass
ratio $\kappa$.
Why merely a Lorentz-force curve?  Why do we not expect, in the
equation for this curve, an additional term involving
the spin of the particle interacting with the Riemann tensor of
the space-time $(M, g_{ab})$?  The reason is the following.
The parameters $m$ and $e$ that appear in the Dirac equation
are in geometrized units --- related to the physical mass and
charge by factors involving Planck's constant $\hbar$.
In order to generate wave packets that follow curves, we must
have $e, m \rightarrow \infty$, and this, for fixed physical
charge and mass, requires $\hbar \rightarrow 0$.  Thus, we
are contemplating a limit in which the spin of the particle
(relative to its mass), and so the effect of that spin on the
particle's motion, goes to zero.  

Strangely enough, there is a different notion of tracking
available in the Dirac case.  Let, say, $e > 0$, so the
charge-current $J^a$ is future-directed causal.  Let us say that a
collection of Dirac solutions tracks, in this new sense, curve $\gamma$
provided:  Given any test field ${\textsc x}_a$, past-directed
causal in a neighborhood of $\gamma$ and timelike at some point
of $\gamma$, there is a field in this collection with
${\bf J}^a\{{\textsc x}\} > 0$.  It turns out that the proof of
Theorem 3 goes through (in fact, somewhat more easily) with this new
notion of tracking:  It follows
that there is some sequence from ${\cal C}$ with the $J^a$
converging, up to rescaling, to $u^a \bm\delta_\gamma$.   Unfortunately,
we cannot conclude from this that $\gamma$ must be a geodesic.

Are there similar results regarding the tracking behavior of solutions
of general symmetric-hyperbolic systems of equations?

\begin{center}{\bf \large Appendix A:  Distributions}\end{center}
\vspace{0.5cm}

We review briefly a few facts about distributions.

Fix, once and for all, a smooth manifold $M$.  By a {\em test field} 
on $M$, we mean a smooth tensor density (of weight 1 -- See Appendix B) 
of compact support.  The test fields, of
given index structure, say ${\textsc t}^b{}_{ac}$, form a vector space.  
A (tensor) {\em distribution}, ${\bf d}_b{}^{ac}$, is a linear map, 
from this vector space to the reals, that
is continuous in the following sense:  Given any $\epsilon > 0$ 
and compact subset $C$ of $M$, there exists an integer $n\geq 0$
and a number $\delta > 0$
such that $|{\bf d}\{{\textsc t}\}| \leq \epsilon$ for every test 
field ${\textsc t}$ whose
support lies in $C$ and which, together with its first $n$
derivatives, is everywhere less than or equal to $\delta$.  In this 
definition, the ``sizes" of the test field and its derivatives are
measured using any fixed positive-definite metric and any derivative
operator on $M$ (the resulting notion of ``continuous" being, 
of course, independent of these choices).  Here, and
hereafter, we use ``$\{\ \}$" to denote the action of a
distribution on a test field.  Distributions with other index 
structures are defined similarly.  Indeed, distributions can
be defined for any kind of fields on $M$, provided only that
those fields have vector-space structure at each point.  For
example, there are distributions based on spinor fields, on charged
fields, on densities, etc.   In the case in which there is given a 
fixed metric,
$g_{ab}$, on the manifold $M$, then we may represent test fields
on $M$ by ordinary tensor fields, converting these to densities
using the alternating tensor of $g_{ab}$.

Note that every smooth tensor field gives rise to a distribution,
where the action of the map is given by contracting that field
with the test field and integrating over $M$.  In fact, every merely
continuous (or even somewhat less well-behaved) tensor field gives
rise to a distribution.  

We can add two distributions (having the same index structure), 
contract a distribution, and take the outer product of a
distribution with a smooth tensor field.  For example, the outer
product of distribution ${\bf d}_b{}^{ac}$ and smooth vector field
$w^d$ is that distribution which, applied to test field 
${\textsc t}_d{}^b{}_{ac}$, yields the number 
${\bf d}\{w^d {\textsc t}_d{}^b{}_{ac}\}$. These operations are
indicated in the usual way (repeated indices for contraction,
juxtaposition for outer product).  Furthermore, given 
any derivative operator, $\nabla_a$, on $M$, that operator can be
extended to act on distributions: For ${\bf d}_b{}^{ac}$ a 
distribution, $\nabla_m {\bf d}_b^{ac}$ is defined as that distribution which,
applied to test field ${\textsc t}^{mb}{}_{ac}$, yields the number
${\bf d}_b{}^{ac}\{-\nabla_m {\textsc t}^{mb}{}_{ac}\}$.  (This is the formula 
suggested by ``integration by parts".)  Note that {\em every}
distribution, no matter how badly behaved, is (infinitely) ``differentiable".
Applied to smooth
tensor fields, regarded as distributions, these operations
reduce to the usual algebraic and differential operations on
those tensor fields.  And, applied to distributions quite generally, 
these operations satisfy all the usual properties. 
So, for example, given any system of linear partial differential 
equation on some tensor fields, there is a corresponding 
system of linear partial differential 
equation on the corresponding distributions.  Thus, we have
a version of Maxwell's equations with distributional Maxwell
field and charge-current.

The {\em support} of a distribution ${\bf d}$ is the smallest closed set 
$C \subset M$ such that ${\bf d}\{{\textsc t}\} = 0$ for 
every test field ${\textsc t}$ whose 
support does not intersect $C$.  So, for example:  The support of 
the derivative of a distribution is a subset of the support of that
distribution.  We say that a distribution is of {\em order $n$} 
(where $n$ is a nonnegative integer) provided the action of ${\bf d}$ can 
be extended, from ($C^\infty$) test fields (of compact support), to
$C^n$ fields of compact support, continuous in the obvious topology.
Thus, every distribution of order $n$ is automatically of every
order $\geq n$; and the derivative of such a distribution is 
automaticaly of order $n+1$.   Note that the outer product of an
order-$n$ distribution and a $C^n$ tensor field makes sense, and
is itself an order-$n$ distribution.
Every distribution of compact support (but not necessarily those
of non-compact support) has some (finite) order.  Note 
that a zero-order (but not in general a higher-order) distribution
annihilates every test field that vanishes on its support.  Every
distribution arising from a smooth tensor field has order
zero, and support given by the support of that tensor field.

As an example, let $(M, g_{ab})$ be a space-time, and $\gamma$ a smooth
timelike curve therein.  Denote by $\bm\delta_\gamma$ the scalar
distribution that assigns, to test field ${\textsc t}$, the result of
first converting ${\textsc t}$ to a scalar field (using the alternating tensor
of $g_{ab}$) and then integrating (with respect to length) over
$\gamma$.  This $\bm\delta_\gamma$ is called the {\em 
delta-distribution} of $\gamma$.  It has order zero and support given
by the curve itself.
It further satisfies $\nabla_a(\bm\delta_\gamma u^a) = 0$, where $\nabla_a$
is the derivative operator determined by $g_{ab}$, and $u^a$ is 
any smooth vector field that, on $\gamma$, is the unit tangent to that 
curve.  Indeed, $\bm\delta_\gamma$ is, up to a constant factor, the
unique order-zero distribution, with support on $\gamma$, with 
this property.  The following 
is a useful fact:  For $\alpha$ any smooth tensor field on $M$, 
the distribution $\nabla_a(\alpha u^a \bm\delta_\gamma)$
depends only on the values of $\alpha$ at the points of $\gamma$,
while the distributions $\alpha u^a \nabla_a \bm\delta_\gamma$ 
and $u^a\nabla_a (\alpha \bm\delta_\gamma)$ do not in
general have this property.  Note that there is no analogous
``delta distribution" for a general null curve.

There is a natural topology on the vector space of distributions of
fixed index structure.  To specify a neighborhood of a distribution
${\bf d}$, fix a finite list, ${\textsc t}_1, {\textsc t}_2, 
\cdots, {\textsc t}_n$, of test fields,
and a number $\epsilon > 0$.  Then the neighborhood consists of all
distributions ${\bf d}'$ such that $|{\bf d}\{{\textsc t}_i\} - 
{\bf d}'\{{\textsc t}_i\}| \leq \epsilon$
for all $i = 1, 2, \cdots, n$.  Thus, for example, if $p_i$ is a
sequence of points of $M$, converging to $p\in M$, then $\bm\delta_{p_i}$,
the sequence of delta-distributions of those points, converges in
this topology to $\bm\delta_p$.  Furthermore:
If a sequence of distributions ${\bf d}_1, {\bf d}_2, \cdots$ converges  
to some distribution
${\bf d}$, then the sequence $\nabla_a {\bf d}_1, \nabla_a {\bf d}_2, \cdots$ 
converges
to $\nabla_a {\bf d}$.  We remark that every distribution on $M$ is a limit,
in this topology, of a sequence of distributions that arise from
smooth tensor fields on $M$.

\begin{center}{\bf \large Appendix B:  Tensor Densities}\end{center}
\vspace{0.5cm}

The test fields on which distributions act are densities.  In the
present context, we wish to impose on these test fields certain inequalities,
such as a version of an energy condition.  But for densities as 
normally defined,
inequalities of this sort make no sense, because of the ambiguity as
to the sign of the alternating tensor.  For present purposes, 
therefore, it is convenient to introduce a slightly different notion 
of a density.

Fix a smooth (say, 4-dimensional) manifold $M$, not necessarily
orientable.  Fix also a point $p$ of $M$.  A {\em tensor density} 
(for order 1) at $p$ is a pair, $(t, \epsilon)$, where $t$ is
a tensor at $p$ and $\epsilon$ is a nonzero, fourth-rank, covariant,
totally antisymmetric tensor at $p$; and where we identify each such
pair with all other pairs of the form $(|a| t, a^{-1} \epsilon)$, as
$a$ runs over the nonzero reals.  A (smooth) density field is
a smooth assignment of a density (of given index structure 
for $t$) to each point of $M$.

Note that a density, as here defined, is slightly different
from what is usually called a ``density".  For example, it makes
sense to say that a scalar density, under the present definition, 
is ``$\geq 0$" (for the demand that $t\geq 0$ is preserved under
$(t,\epsilon) \rightarrow (|a|t, a^{-1}\epsilon)$).  Note that every 
manifold $M$, even a non-orientable one, admits some positive scalar 
density field.  

We can add two density fields having the same index structure (by
choosing representatives at each point having the same underlying 
$\epsilon$, and then adding the tensors $t$ of those representatives).
We can also contract a density field, and take the outer product
of a density field and a tensor field, in the obvious way.  
These algebraic operations have all the standard properties.
Thus, for example, the density fields, of given index structure,
form a vector space.  Finally, given any 
derivative operator, $\nabla_a$, on tensor fields on $M$, we
can extend that operator to act also on density fields, the result
of this operation again being a density field.  And
this operator on density fields has all the usual properties,
such as linearity and the Leibnitz rule.

Let $\alpha$ be a smooth scalar density field on $M$, and let
$O\subset M$ be open.  Then $\int_O \alpha$ makes sense (provided
the integral converges). Indeed, if $O$ is orientable, choose any
alternating tensor field on $O$, and use it to convert $\alpha$ into
an ordinary scalar field and also as the volume element to integrate
that scalar over $O$.  The result is independent of the choice
of alternating tensor.  If $O$ is non-orientable, write it as a
union of orientable regions, and define the integral over $O$ as
the sum of the integrals over these regions.  Note that this integral,
so defined, does {\em not} require that the manifold $M$ be orientable, 
nor that there be specified any volume-element on $M$.  As an
example, we have:  If $\alpha \leq 0$, then $\int_O \alpha \leq 0$.

Finally, consider the case in which there is specified some
fixed Lorentz metric $g_{ab}$ on $M$ (which, still, may be 
non-orientable).  Then this $g_{ab}$ gives rise to an alternating 
tensor, $\epsilon_{abcd}$, at each point of $M$,
up to sign.  We may now use this $\pm \epsilon_{abcd}$ to convert
each tensor density field on $M$ into an ordinary tensor field
(of the same index structure).  Note that a nonnegative scalar
density is thereby converted to a nonnegative scalar field.

\end{document}